\documentclass[twocolumn,amsmath,amssymb,a4paper,prb,superscriptaddress]{revtex4-1}
\usepackage{graphicx}
\usepackage{natbib}
\usepackage{multirow}
\usepackage{amsmath}
\usepackage{url}
\usepackage{lineno}
\usepackage{hyperref}
\usepackage{makecell}
\begin{document}

\title{\texttt{Spglib}: a software library for crystal symmetry search}

\author{Atsushi Togo}
\email{togo.atsushi@nims.go.jp}
\affiliation{Center for Basic Research on Materials, National Institute for
  Materials Science, Tsukuba, Ibaraki 305-0047, Japan}
\affiliation{Center for
  Elements Strategy Initiative for Structural Materials, Kyoto University, Sakyo,
  Kyoto 606-8501, Japan}

\author{Kohei Shinohara}
\affiliation{Preferred Networks, Inc., Tokyo 100-0004, Japan.}

\author{Isao Tanaka}
\affiliation{Center for Elements Strategy Initiative for Structural
  Materials, Kyoto University, Sakyo, Kyoto 606-8501, Japan}
\affiliation{Department of Materials Science and
  Engineering, Kyoto University, Sakyo, Kyoto 606-8501, Japan}
\affiliation{Nanostructures Research Laboratory, Japan Fine Ceramics
  Center, Atsuta, Nagoya 456-8587, Japan}

\begin{abstract}
  A computer algorithm to search symmetries of crystal structures as implemented
  in the \texttt{spglib} code is described. An iterative algorithm is employed
  to robustly identify space group types tolerating a certain amount of distortion
  in the crystal structures. The source code is distributed under the 3-Clause
  BSD License, a permissive open-source software license. This paper
  focuses on the algorithm for identifying the space group symmetry of the crystal
  structures.
\end{abstract}

\maketitle

% \tableofcontents

\section{Introduction}

\begin{figure}[ht]
  \begin{center}
    \includegraphics[height=0.82\textheight]{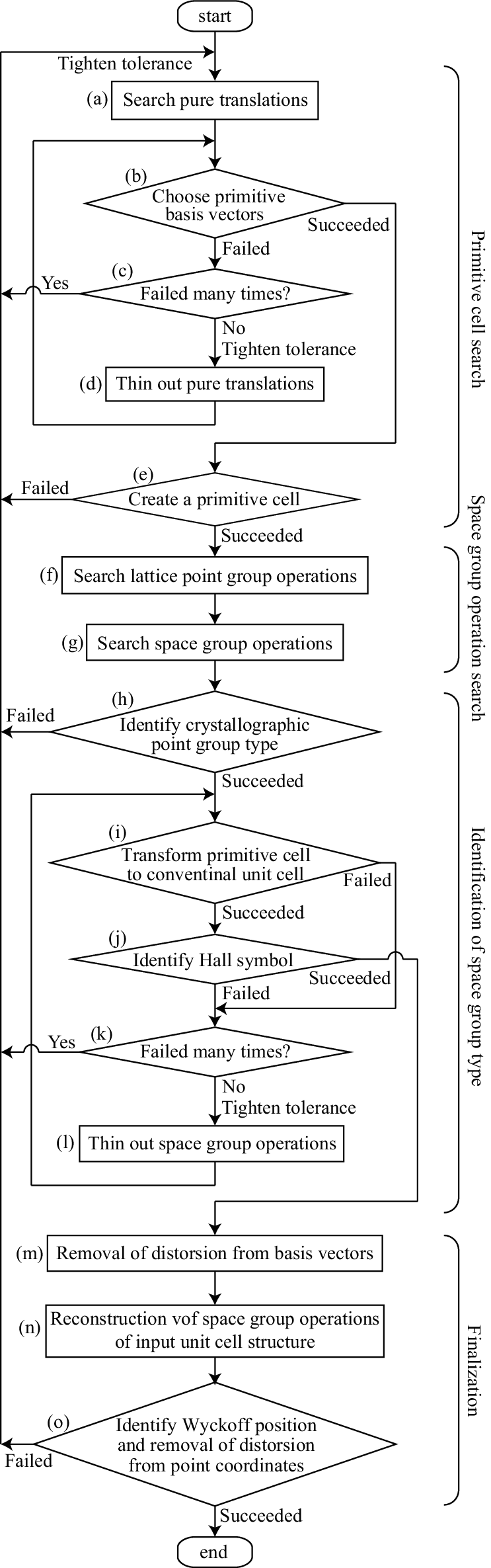}
    \caption{\label{fig:flowchart}
      Flowchart of algorithm in the \texttt{spglib} code.}
  \end{center}
\end{figure}

Crystal symmetry is essential information to understand various crystal
properties. It is also useful to compress information of physical states of
crystals, e.g., in electronic structure calculations.
The crystal symmetry is composed of a set of symmetry operations that
map a crystal structure onto itself, and the set forms a space group. There are
230 space group types. Each crystal structure is uniquely assigned to one of the
space-group types.\cite{ITA} Here, the crystal structure is a mathematical
model, where equivalent unit cells are arranged on a periodic lattice, and the
unit cell is defined by basis vectors and point coordinates of atoms labeled by
atomic species.

Assuming the periodicity, a crystal structure is represented by numerical values
of the basis vectors and point coordinates in a unit cell. The symmetry
operations are also represented numerically with respect to the unit cell. Due
to the periodicity of unit cells, another linear combination of the basis
vectors can be also a valid unit cell, which results in different matrix
representations of the symmetry operations. It is inconvenient to represent the
same crystal symmetry in different representations. To obtain a unique matrix
representation of symmetry operations, conventional choices and settings of unit
cell representations are employed in crystallography. These matrix
representations of the space-group types are tabulated in {\it International
    Tables for Crystallography Volume A} (ITA).\cite{ITA}

Given a crystal structure, its symmetry operations are searched computationally,
then, the space-group type is identified. If the crystal structure is provided
in a standard setting, the identification is straightforward. However, it is
often provided in a different choice of basis vectors and with arbitrary origin
shift, i.e., non-standard setting. Furthermore, when the basis vectors and point
coordinates are slightly distorted with respect to expected crystal symmetry,
the symmetry search becomes a challenging task. The \texttt{spglib} code has been
developed to solve these difficulties.

There already exist crystal symmetry finding codes. The \texttt{FINDSYM} code in
the ISOTROPY Software Suite~\cite{Stokes-2005, Findsym, ISOTROPY} has been well
known, however, the source code is unavailable in public. The
\texttt{cctbx}~\cite{Grosse-Kunstleve-cctbx-2002, Grosse-Kunstleve-Niggli-2004,
  Grosse-Kunstleve-II-2002, Grosse-Kunstleve-I-1999, cctbx} code is an open-source
computational crystallography toolbox under a BSD-type variant license. The
\texttt{sginfo}~\cite{Grosse-Kunstleve-cctbx-2002, Grosse-Kunstleve-Niggli-2004,
  Grosse-Kunstleve-II-2002, Grosse-Kunstleve-I-1999} code, currently distributed
under an open source software license, has been superseded by the space group
toolbox (\texttt{sgtbx}) in the \texttt{cctbx} code. The
\texttt{AFLOW-SYM}~\cite{AFLOW-SYM} code is an open-source crystal symmetry
analysis code under the GNU general public license. The \texttt{spglib} code is
another open-source code distributed under the 3-clause BSD license that is a
permissive open-source software license.\cite{spglib}

Using the \texttt{spglib} code, space group operations are searched from a
crystal structure in which small deviations of atomic positions from their ideal
positions should be tolerated. As a result, we obtain the coset
representatives of the space group $\mathbb{S}$ with respect to the translation
group $\mathbb{T}$. Some other useful information about the crystal symmetry is
derived simultaneously with running the algorithm. After many years of
development, the \texttt{spglib} code has become one of the key software
libraries in the scientific community. Since many software packages rely on it
already, the \texttt{spglib} code is expected to be well maintained and live
longer. Therefore, this paper aims to invite developers of the
\texttt{spglib} code by presenting a detailed algorithm implemented in the
\texttt{spglib} code.

In the algorithm outlined in this paper, we employ an iterative approach to
determine space group operations for the initial input crystal structure at a
given distance tolerance. The flowchart of the algorithm is presented in
Fig.~\ref{fig:flowchart}. There are four stages in the algorithm: primitive cell
search, space group operation search, identification of space group type, and
finalization. These stages are subdivided into smaller steps. The algorithm
continues until the obtained symmetry operations satisfy all given crystallographic
constraints under minimum adjustment of the tolerance value. There are nested
iteration loops in the algorithm. Some loops are explicitly depicted in
Fig.~\ref{fig:flowchart}. A few more loops are contained inside the blocks. The
tolerance value is decreased when the operation fails in each loop. One
exception is step (e), where both increasing and decreasing the tolerance
value are attempted.

This paper concentrates on the algorithms designed for searching space group
operations and identifying the space group types. The algorithm related to the
magnetic space group, which builds upon the space group algorithm, is discussed
in another publication.\cite{Shinohara-MSG} A tolerance parameter, which is
crucial for the algorithm in the \texttt{spglib} code, is detailed in the
following section. Notations, algebra, and look-up tables utilized in this paper
are compiled in the Appendices.

\section{Tolerance parameter in crystal symmetry operation search}
\label{sec:sym_search}

In the \texttt{spglib} code, numerical searches for symmetry operations of a
provided input unit cell utilize a small Euclidean distance $\epsilon$ as a
tolerance parameter. The value of the tolerance parameter is adjusted in the
process of the algorithm to identify a possible space group. This section
explains how it is employed to examine the symmetrical equivalence of two atomic
points. How it is adjusted is explained in the following specific sections.
Symbols used in this section  are summarized in
Appendices~\ref{sec:basis-vectors}--\ref{sec:sp-operation}.

\subsection{Lattice translation to a point near origin}

Each element of point coordinates, $\boldsymbol{x}=(x_1, x_2, x_3)^\top$,
is confined within the interval $[-0.5,0.5)$ by the operation:
\begin{equation}
  \label{eq:rem1}
  x_i \leftarrow x_i - \lfloor x_i\rceil, \; i=1, 2, 3,
\end{equation}
where $\lfloor x_i \rceil$ denotes rounding $x_i$ to the nearest integer. For
the coordinates, this operation is concisely represented as
\begin{equation}
  \label{eq:rem1-cood}
  \boldsymbol{x} \leftarrow \boldsymbol{x} - \lfloor \boldsymbol{x} \rceil,
\end{equation}
and similarly for vectors,
\begin{equation}
  \label{eq:rem1-vec}
  \boldsymbol{w} \leftarrow \boldsymbol{w} - \lfloor \boldsymbol{w} \rceil.
\end{equation}

\subsection{Determination of equivalent points under lattice periodicity}
The tolerance parameter $\epsilon$ is used to identify if two points
$\tilde{\boldsymbol{x}}$ and $\boldsymbol{x}'$ occupy the same atomic site or
not. Using the difference of positions $\Delta\boldsymbol{x} =
  \tilde{\boldsymbol{x}} - \boldsymbol{x}'$, this is examined by
\begin{equation}
  \label{eq:Euclidean-dist}
  \left|
  (\mathbf{a}, \mathbf{b}, \mathbf{c})\Delta\boldsymbol{x}
  \right| < \epsilon,
\end{equation}
where $(\mathbf{a}, \mathbf{b}, \mathbf{c})$ represent the basis vectors of the
unit cell. To accommodate the periodicity of the lattice, the condition
(\ref{eq:Euclidean-dist}) is reformulated using the operation (\ref{eq:rem1-vec}) as
\begin{equation}
  \label{eq:red-Euclidean-dist}
  \left| ( \mathbf{a}, \mathbf{b}, \mathbf{c})(\Delta\boldsymbol{x} -
  \lfloor\Delta\boldsymbol{x}\rceil ) \right| < \epsilon.
\end{equation}
This expression is frequently used in the implementation.

\subsection{Examination of a symmetry operation}
Given a space group operation $(\boldsymbol{W},\boldsymbol{w})$, which may or
may not be a valid symmetry operation, an atomic point $\boldsymbol{x}$ is
transformed to $\tilde{\boldsymbol{x}} = (\boldsymbol{W},\boldsymbol{w})
\boldsymbol{x}$. If $(\boldsymbol{W},\boldsymbol{w})$ represents a valid space
group operation, $\tilde{\boldsymbol{x}}$ must be located at one of atomic sites
$\boldsymbol{x}'$ with the same atomic type. In the algorithm, this is examined
for all atoms in the input unit cell or the primitive cell using
condition~(\ref{eq:red-Euclidean-dist}). When this is satisfied, the given space
group operation $(\boldsymbol{W},\boldsymbol{w})$ is accepted.

\section{Machine precision issue}
\label{sec:machine-precision}

It is assumed that the \texttt{spglib} code is used on 32- and 64-bit computer
systems. As reported in Ref.~\onlinecite{Grosse-Kunstleve-Niggli-2004},
comparisons of lengths have to be carefully implemented in the code due to
finite machine precision. For example, an inequality $x < y$ may be implemented
in $x < y - \epsilon'$ with a small positive value
$\epsilon'$.\cite{Grosse-Kunstleve-Niggli-2004} Most of the inequalities that
appear in the implementation are in the style of
Eq.~(\ref{eq:red-Euclidean-dist}) and none of the special care is applied for
this case. On the other hand, the operation to take modulo by $\mathbb{Z}$ is
performed using the above inequality using $\epsilon'$. With this operation, a
value $x_i$ is converted to fall within the interval $[-\epsilon',
1-\epsilon')$. $\epsilon'=10^{-10}$ is employed in the current version of the
\texttt{spglib} code.

\section{Primitive cell search}
\label{sec:primitive-cell-search}
In the first stage of the space group symmetry search, a primitive cell is
determined from lattice points in the input unit cell. The lattice points are
obtained through pure translation operation search.

\subsection{Step (a): Searching pure translation operations}
The input unit cell contains multiple lattice points in it if it is not a
primitive cell. These lattice points are obtained as translation parts
$\boldsymbol{w}_{\boldsymbol{I}}$ of pure translation operations of
$\{(\boldsymbol{I},\boldsymbol{w}_{\boldsymbol{I}})\}$ in the input unit cell,
where $\boldsymbol{I}$ is the identity matrix. The pure translation operations
of $\{(\boldsymbol{I},\boldsymbol{w}_{\boldsymbol{I}})\}$ are searched as
follows. Candidates of the translation parts
$\boldsymbol{w}_{\boldsymbol{I}}^\text{c}$ are selected from vectors that extend
from a fixed atomic site $\boldsymbol{x}$ to all atomic sites $\boldsymbol{x}'$
of the same atomic type $A$, i.e., $\boldsymbol{w}_{\boldsymbol{I}}^\text{c} =
  \boldsymbol{x}'_A - \boldsymbol{x}_A$. To minimize computational demand, the
fixed atomic site is chosen among atoms having an atomic type that comprises the
smallest number of atoms.\cite{Hannemann-1998} Each candidate vector
$\boldsymbol{w}_{\boldsymbol{I}}^\text{c}$ is examined as described in
Sec.~\ref{sec:sym_search}. If all $\Delta \boldsymbol{x}_A = (\boldsymbol{I},
  \boldsymbol{w}_{\boldsymbol{I}}^\text{c})\boldsymbol{x}_A - \boldsymbol{x}'_A$
for all $\boldsymbol{x}_A$ satisfy the condition~(\ref{eq:red-Euclidean-dist}),
this $(\boldsymbol{I},\boldsymbol{w}_{\boldsymbol{I}}^\text{c})$ is a pure
translation operation.

If the input unit cell is a primitive cell, only one
$(\boldsymbol{I},\boldsymbol{w}_{\boldsymbol{I}})$ with
$\boldsymbol{w}_{\boldsymbol{I}} = (0, 0, 0)^\top$ should be found,
otherwise a set of multiple pure translation operations of
$\{(\boldsymbol{I},\boldsymbol{w}_{\boldsymbol{I}})\}$ are obtained.

In typical use cases, this step is the most computationally demanding part in
the entire process. The brute-force algorithm has a time complexity of
$\mathcal{O}(N^3)$, with $N$ denoting the number of type $A$ atoms.
However, this complexity is empirically reduced to $\mathcal{O}(N^2\log{N})$ by
sorting the atoms, which minimizes the worst-case
scenarios.\cite{Lamparski-thesis}

\subsection{Step (b): Choosing basis vectors of primitive cell}
Candidates for three basis vectors of a primitive cell
$(\mathbf{a}_{\text{p}}^{\text{c}}, \mathbf{b}_{\text{p}}^{\text{c}},
  \mathbf{c}_{\text{p}}^{\text{c}})$ are chosen from the set of vectors
$\mathbf{T}_\text{i} \cup \mathbf{T}_\text{p}$, where $\mathbf{T}_\text{i} =
  \{\mathbf{a}_\text{i}, \mathbf{b}_\text{i}, \mathbf{c}_\text{i}\}$ is the set
composed of the basis vectors of the input unit cell, and $\mathbf{T}_\text{p} =
  \{\boldsymbol{w}_{\boldsymbol{I}}\}$ found at step (a). The three basis vectors
are chosen to create a right-handed coordinate system. The volume of the primitive
cell, $V_\text{p}$, is expected to be approximately the volume of the input
unit cell, $V_\text{i}$, divided by $|\mathbf{T}_\text{p}|$. Therefore, the
basis vectors of the primitive cell are searched under the condition:
\begin{equation}
  \label{eq:prim-basis-cond}
  |\mathbf{T}_\text{p}| = \lfloor V_\text{i} / V_\text{p} \rceil.
\end{equation}

\subsection{Step (c): Failure of finding primitive cell basis vectors}
For distorted input unit cells, the condition~(\ref{eq:prim-basis-cond}) may not
always be satisfied. For example, the number of the pure translations found can
either be more than or less than those expected.  In this case, the sequence of
steps (b), (c), and (d) is iterated by reducing the tolerance value. If this
loop is repeated many times, the procedure restarts from step (a) with the
tolerance value reduced from that previously used in step (a).

\subsection{Step (d): Thinning out pure translations}
Some of the pure translations  that do not satisfy the
condition~(\ref{eq:prim-basis-cond}) are discarded by re-examining the existing
pure translations with a tightened tolerance value. Typically, this operation is
far more computationally efficient than restarting from step (a) with a
tightened tolerance value.

\subsection{Step (e): Creating a primitive cell}
The primitive cell basis vectors $(\mathbf{a}_{\text{p}}^{\text{c}},
  \mathbf{b}_{\text{p}}^{\text{c}}, \mathbf{c}_{\text{p}}^{\text{c}})$ found at
step (b) are transformed to a different set of primitive cell basis vectors
$(\mathbf{a}_\text{p}, \mathbf{b}_\text{p}, \mathbf{c}_\text{p})$ by the
Delaunay reduction.\cite{Delaunay1933, ITA} This transformation is written as
\begin{equation}
  \label{eq:primitive-basis-vectors}
  (\mathbf{a}_\text{p}, \mathbf{b}_\text{p}, \mathbf{c}_\text{p}) =
  (\mathbf{a}_{\text{p}}^{\text{c}},
  \mathbf{b}_{\text{p}}^{\text{c}}, \mathbf{c}_{\text{p}}^{\text{c}})
  \mathbf{Q}_\text{D}.
\end{equation}
$\mathbf{Q}_\text{D}$ is an integer matrix and is chosen such that
$\det(\mathbf{Q}_\text{D}) = 1$.

Similarly, the transformation of the basis vectors of the primitive cell basis
vectors to those of the input unit cell $(\mathbf{a}_\text{i},
  \mathbf{b}_\text{i}, \mathbf{c}_\text{i})$ in a right handed coordinate system
is written by the change-of-basis matrix
$\mathbf{Q}_{\text{p}\rightarrow\text{i}}$ as
\begin{equation}
  \label{eq:p2i-tmat}
  (\mathbf{a}_\text{i}, \mathbf{b}_\text{i}, \mathbf{c}_\text{i}) \approx
  (\mathbf{a}_\text{p}, \mathbf{b}_\text{p}, \mathbf{c}_\text{p}) \mathbf{Q}_{\text{p}\rightarrow\text{i}}
\end{equation}
where $\mathbf{Q}_{\text{p}\rightarrow\text{i}}$ is chosen as an integer matrix
with $\det(\mathbf{Q}_{\text{p}\rightarrow\text{i}}) \geq 1$ although
Eq.~(\ref{eq:p2i-tmat}) is an approximation if $\mathbf{T}_\text{p} =
  \{\boldsymbol{w}\}$ found at step (a) is distorted with respect to the lattice
of the primitive cell.

Point coordinates in the input unit cell, $\boldsymbol{x}_\text{i}$,
are transformed to their corresponding coordinates in the primitive cell,
$\boldsymbol{x}_{\mathrm{p}*}$, by
\begin{equation}
  \boldsymbol{x}_{\mathrm{p}*} = \mathbf{Q}_{\text{p}\rightarrow\text{i}} \boldsymbol{x}_\text{i}.
\end{equation}
where $\boldsymbol{x}_{\mathrm{p}*}$ is brought in the interval $[-0.5,0.5)$ by
the operation (\ref{eq:rem1}). When
$\det(\mathbf{Q}_{\text{p}\rightarrow\text{i}}) > 1$, multiple
$\boldsymbol{x}_\text{i}$ should be mapped to a point
$\boldsymbol{x}_{\mathrm{p}*}$ with respect to $(\mathbf{a}_\text{p},
  \mathbf{b}_\text{p}, \mathbf{c}_\text{p})$ using
condition~(\ref{eq:red-Euclidean-dist}). The tolerance value $\epsilon$ in
condition~(\ref{eq:red-Euclidean-dist}) is adjusted until the multiplicity
becomes $\det(\mathbf{Q}_{\text{p}\rightarrow\text{i}})$ for all translationally
independent $\boldsymbol{x}_{\text{p}*}$. If this fails, the procedure restarts
from step (a) with the tolerance value tightened from that previously used at
step (a).

Upon successful verification, the point coordinates of translationally
equivalent atoms are averaged to produce $\boldsymbol{x}_\text{p}$ along with
boundary treatment for $\boldsymbol{x}_{\mathrm{p}*}$ located close to the
boundary of $[-0.5,0.5)$. Finally, a primitive cell is created from
$(\mathbf{a}_\text{p}, \mathbf{b}_\text{p}, \mathbf{c}_\text{p})$ and
$\{\boldsymbol{x}_\text{p}\}$.

\section{Space group operation search}
\label{sec:spg-search}

The purpose of the second stage is to search a set of space group operations
$\{(\boldsymbol{W}_{\text{p}},\boldsymbol{w}_{\text{p}})\}$ of the primitive
cell obtained in the first stage. Candidates of the rotation parts
$\{\boldsymbol{W}^\mathrm{c}\}$ are given by an exhaustive search of lattice
point group operations. Using obtained $\{\boldsymbol{W}^\mathrm{c}\}$ and the
point coordinates $\{\boldsymbol{x}_\text{p}\}$, a set of the space group
operations $\{(\boldsymbol{W}_{\text{p}},\boldsymbol{w}_{\text{p}})\}$ is
searched.

\subsection{Step (f): Searching lattice point group operations}

A set of possible candidates of lattice point group operations
$\{\boldsymbol{W}^\mathrm{c}_\mathrm{L}\}$ is exhaustively generated by filling
the matrix elements with -1, 0, or 1 under the constraint of
$|\det(\boldsymbol{W}^\mathrm{c}_\mathrm{L})|=1$. As described in
Appendix~\ref{sec:pg-lattice}, a set of lattice point group operations of the
primitive cell, $\{\boldsymbol{W}_\mathrm{L}\}$, is searched within
$\{\boldsymbol{W}^\mathrm{c}_\mathrm{L}\}$ using the metric tensor
$\boldsymbol{G}$ that is rotated by $\boldsymbol{W}_\mathrm{L}$ as
$\tilde{\boldsymbol{G}}=(\boldsymbol{W}_\mathrm{L})^\top \boldsymbol{G}
  \boldsymbol{W}_\mathrm{L}$.

Comparison of $\boldsymbol{G}$ and $\tilde{\boldsymbol{G}}$ is performed as
follows. The diagonal elements of the matrix $\boldsymbol{G}$ provide
information about the lengths of the basis vectors, while the off-diagonal
elements indicate the angles between these vectors. The differences in lengths
can be straightforwardly compared using the tolerance $\epsilon$ based on
Euclidean distance. For angle comparison, an  angle tolerance parameter can be
employed. However, this addition of an extra tolerance parameter may complicate
the usage. Therefore, in the current implementation, if an angle tolerance is
not explicitly specified, the distance tolerance $\epsilon$ is approximated for
angle comparisons. This approach is applied, for instance, in the evaluation of
$G_{12}$, as shown in the following equation:
\begin{equation}
  \label{eq:angle-tolerance}
  \sin \left| \Delta \theta \right|
  \sqrt{\frac{(|\tilde{\mathbf{a}}| + |\mathbf{a}|) (|\tilde{\mathbf{b}}| + |\mathbf{b}|)}{4}}
  < \epsilon,
\end{equation}
where $\Delta \theta$ is the angle difference between the two pairs of vectors
$\mathbf{a}$--$\mathbf{b}$ and $\tilde{\mathbf{a}}$--$\tilde{\mathbf{b}}$,
\begin{equation}
  \Delta \theta =
    \arccos \left(
      \frac{\tilde{\mathbf{a}}\cdot\tilde{\mathbf{b}}}{|\tilde{\mathbf{a}}| |\tilde{\mathbf{b}}|}
    \right)
    - \arccos\left(
      \frac{\mathbf{a}\cdot\mathbf{b}}{|\mathbf{a}| |\mathbf{b}|}
    \right).
\end{equation}
The angle difference $\Delta \theta$ is compared with the tolerance value with
the averaged lengths of basis vectors.
The left-hand side of Eq.~\eqref{eq:angle-tolerance} is given by the matrix
elements of $\boldsymbol{G}$ and $\tilde{\boldsymbol{G}}$ such as
$\mathbf{a}\cdot\mathbf{b} = \sqrt{G_{12}}$, $|\mathbf{a}| =
  \sqrt{G_{11}}$, and $|\mathbf{b}| = \sqrt{G_{22}}$.

In a similar way
to that applied in step (a), the corresponding translation part
$\boldsymbol{w}_{\text{p}}$ is searched using $\boldsymbol{W}_\mathrm{L}$
instead of $\boldsymbol{I}$ in step (a).

\subsection{Step (g): Searching space group operations}
\label{sec:searching-space-group-operations}
A set of the space group operations
$\{(\boldsymbol{W}_{\text{p}},\boldsymbol{w}_{\text{p}})\}$ is searched in the
following way. The rotation matrices $\{\boldsymbol{W}_\mathrm{L}\}$ found in
step (f) are used as candidates of the rotation parts of
$\{(\boldsymbol{W}_{\text{p}},\boldsymbol{w}_{\text{p}})\}$. From
Eq.~(\ref{eq:spg-operation}), a space group operation
$(\boldsymbol{W}_{\text{p}},\boldsymbol{w}_{\text{p}})$ satisfies
$\tilde{\boldsymbol{x}} = \boldsymbol{W}_{\text{p}}\boldsymbol{x} +
  \boldsymbol{w}_{\text{p}}$. Therefore, candidates of translation parts for
$\boldsymbol{W}_\mathrm{L}$ are given by $\boldsymbol{w}_\mathrm{p}^\mathrm{c}
  = \tilde{\boldsymbol{x}} - \boldsymbol{W}_\mathrm{L}\boldsymbol{x}$ over
possible combinations of $\boldsymbol{x}$ and $\tilde{\boldsymbol{x}}$.

To limit its search space, for a fixed $\boldsymbol{x}$,
$\tilde{\boldsymbol{x}}$ are selected from the all atoms with the same atomic
type as $\boldsymbol{x}$ in the primitive cell. Then,
$(\boldsymbol{W}_\mathrm{L}, \boldsymbol{w}^\mathrm{c}_\mathrm{p})$ is examined
by applying condition~(\ref{eq:red-Euclidean-dist}) with $\Delta \boldsymbol{x}
  = (\boldsymbol{W}_\mathrm{L},
  \boldsymbol{w}^\mathrm{c}_\mathrm{p})\boldsymbol{x} - \tilde{\boldsymbol{x}}$
for all $\boldsymbol{x}$ and $\tilde{\boldsymbol{x}}$. If none of
$\boldsymbol{w}^\mathrm{c}_\mathrm{p}$ is found, this
$\boldsymbol{W}_\mathrm{L}$ is discarded, otherwise $(\boldsymbol{W}_\mathrm{L},
  \boldsymbol{w}^\mathrm{c}_\mathrm{p})$ is adopted as
$(\boldsymbol{W}_{\text{p}},\boldsymbol{w}_{\text{p}})$.  This procedure is
repeated over all elements of $\{\boldsymbol{W}_\mathrm{L}\}$. Finally, a set of
the space group operations
$\{(\boldsymbol{W}_{\text{p}},\boldsymbol{w}_{\text{p}})\}$ is obtained. In the
next stage, it is verified that
$\{(\boldsymbol{W}_{\text{p}},\boldsymbol{w}_{\text{p}})\}$ constitutes the coset
representatives of the factor group $\mathbb{S}/\mathbb{T}$.

\section{Identification of space group type}
In this third stage, a space group type is identified by comparing the set of
the space group operations
$\{(\boldsymbol{W}_{\text{p}},\boldsymbol{w}_{\text{p}})\}$ obtained in the last
stage with those sets coded in the Hall symbols.\cite{Hall-1981,ITB} To achieve
this, the primitive cell is transformed into the corresponding conventional unit
cell in a specific setting. The space group operations for this setting are
matched with those matrix representations decoded from the Hall symbols and the
origin shift is simultaneously determined at the matching
process.\cite{Grosse-Kunstleve-I-1999} The algorithm presented in this section
follows almost exactly as that reported by Grosse-Kunstleve and Adams in
Ref.~\onlinecite{Grosse-Kunstleve-I-1999}, and it is described as implemented in
the \texttt{spglib} code.

\subsection{Step (h): Identify crystallographic point group}

The crystallographic point group is given by collecting rotation parts of the
space group operations, i.e., $\mathbb{P}=\{\boldsymbol{W}_{\text{p}}|
  (\boldsymbol{W}_{\text{p}},\boldsymbol{w}_{\text{p}}) \in
  \{(\boldsymbol{W}_{\text{p}},\boldsymbol{w}_{\text{p}})\}\}$. The
crystallographic point group type is identified from the traces and determinants
of the matrices of $\{\boldsymbol{W}_{\text{p}}\}$ by using the look-up Tables
\ref{tab:rotations} and \ref{tab:crystal-class} presented in
Appendix~\ref{sec:cryst-point-group}. If the identification of the
crystallographic point group type failed, the procedure restarts from step (a)
with the tolerance value tightened from that previously used in step (a).

\subsection{Step (i): Transformation from primitive cell to
  conventional unit cell}

Laue class is the information necessary to transform the basis vectors of the
primitive cell to those of the corresponding conventional unit cell. It is
easily known once the crystal class, which is equivalent to the crystallographic
point group type, is determined as shown in
Table \ref{tab:crystal-class}.

The transformation of the basis vectors of the primitive cell
$(\mathbf{a}_\mathrm{p}, \mathbf{b}_\mathrm{p},\mathbf{c}_\mathrm{p})$
to those of the conventional unit cell $(\mathbf{a}_\mathrm{c},
  \mathbf{b}_\mathrm{c},\mathbf{c}_\mathrm{c})$ is written as
\begin{equation}
  \label{eq:tmat-prim-to-conv}
  (\mathbf{a}_\mathrm{c},
  \mathbf{b}_\mathrm{c},\mathbf{c}_\mathrm{c}) = (\mathbf{a}_\mathrm{p},
  \mathbf{b}_\mathrm{p},\mathbf{c}_\mathrm{p}) \mathbf{M}'.
\end{equation}
$\mathbf{M}'$ is constructed from three axis directions
$(\mathbf{e}_x, \mathbf{e}_y, \mathbf{e}_z)$ such as
\begin{equation}
  \mathbf{M}' = \begin{bmatrix}
    e_{x_1} & e_{y_1} & e_{z_1} \\
    e_{x_2} & e_{y_2} & e_{z_2} \\
    e_{x_3} & e_{y_3} & e_{z_3} \\
  \end{bmatrix},
\end{equation}
with $\det(\mathbf{M}')>0$. These axis directions are determined from
rotation axes that characterize the Laue class. The rotation axis
direction of each $\boldsymbol{W}_{\text{p}}$ is found by solving the following
equation:
\begin{equation}
  \label{eq:axis1}
  \boldsymbol{W}^\mathrm{prop}  \mathbf{e} = \mathbf{e},
\end{equation}
where $\boldsymbol{W}^\mathrm{prop} =
  \det(\boldsymbol{W}_{\text{p}})\boldsymbol{W}_{\text{p}}$ is the proper
  rotation matrix of $\boldsymbol{W}_{\text{p}}$. The rotation axis direction
  $\mathbf{e}$ can be determined through an exhaustive search, wherein three
  integer values are evaluated as the components of $\mathbf{e}$. The rotation
  order $n$ of $\boldsymbol{W}_{\text{p}}$ is defined by the smallest $n>0$ that
  satisfies
\begin{equation}
  (\boldsymbol{W}^\mathrm{prop})^n = \boldsymbol{I}.
\end{equation}

\begin{table}
  \caption{\label{tab:axis-search} Axis directions for Laue classes.
    $n^\mathrm{pri}$, $n^\mathrm{sec}$, and $n^\mathrm{ter}$ are the rotation
    orders of the primary ($\boldsymbol{W}^\mathrm{pri}$), secondary
    ($\boldsymbol{W}^\mathrm{sec}$), and ternary ($\boldsymbol{W}^\mathrm{ter}$)
    proper rotation matrices, respectively.}
  \begin{center}
    \begin{tabular}{lcl}
      \hline
      Laue class  &  & Condition implemented in the \texttt{spglib} code    \\
      \hline
      $\bar{1}$   &  & Do nothing                                           \\
      \hline
      $2/m$       &  & $n^\mathrm{pri}=2$ for $\mathbf{b}_\mathrm{c}$,
      $\mathbf{S}\mathbf{e}^\mathrm{sec}=\mathbf{0}$,
      $\mathbf{S}\mathbf{e}^\mathrm{ter}=\mathbf{0}$,
      $\mathbf{e}^\mathrm{sec} \neq \mathbf{e}^\mathrm{ter}$                \\
      \hline
      $4/m$       &  & $n^\mathrm{pri}=4$ for $\mathbf{c}_\mathrm{c}$,
      $\mathbf{S}\mathbf{e}^\mathrm{sec}=\mathbf{0}$,
      $\mathbf{e}^\mathrm{ter} = \boldsymbol{W}^\mathrm{pri}
      \mathbf{e}^\mathrm{sec}$                                              \\
      $4/mmm$     &  & Same as $4/m$                                        \\
      $\bar{3}$   &  & $n^\mathrm{pri}=3$ for $\mathbf{c}_\mathrm{c}$,
      $\mathbf{S}\mathbf{e}^\mathrm{sec}=\mathbf{0}$,
      $\mathbf{e}^\mathrm{ter} = \boldsymbol{W}^\mathrm{pri}
      \mathbf{e}^\mathrm{sec}$                                              \\
      $\bar{3}m$  &  & Same as $\bar{3}$                                    \\
      $6/m$       &  & Same as $\bar{3}$                                    \\
      $6/mmm$     &  & Same as $\bar{3}$                                    \\
      \hline
      $mmm$       &  & $n^\mathrm{pri}=n^\mathrm{sec}=n^\mathrm{ter}=2$ for
      $\mathbf{a}_\mathrm{c}$,
      $\mathbf{b}_\mathrm{c}$, $\mathbf{c}_\mathrm{c}$                      \\
      $m\bar{3}$  &  & Same as $mmm$                                        \\
      $m\bar{3}m$ &  & $n^\mathrm{pri}=n^\mathrm{sec}=n^\mathrm{ter}=4$ for
      $\mathbf{a}_\mathrm{c}$,
      $\mathbf{b}_\mathrm{c}$, $\mathbf{c}_\mathrm{c}$                      \\
      \hline
    \end{tabular}
  \end{center}
\end{table}

Except for the Laue class $\bar{1}$, the primary axis direction
$\mathbf{e}^\mathrm{pri}$ is determined by selecting a primary proper
rotation matrix $\boldsymbol{W}^\mathrm{pri}$ of the rotation order
$n^\mathrm{pri}$ presented in Table~\ref{tab:axis-search}. The axis
direction $\mathbf{e}'$ perpendicular to $\mathbf{e}^\text{pri}$ is
determined to satisfy the following equation:
\begin{equation}
  \label{eq:axis2}
  \mathbf{S}\mathbf{e}'=\mathbf{0},
\end{equation}
where
$\mathbf{S}=\sum_{i=1}^{n_\text{pri}}\boldsymbol{W}^\text{pri}_{i}$.
The conditions that the primary, secondary ($\mathbf{e}^\mathrm{sec}$),
and ternary ($\mathbf{e}^\mathrm{ter}$) axis directions have to satisfy
are listed in Table~\ref{tab:axis-search} for Laue classes. For the Laue
class $\bar{1}$, $\mathbf{M}'$ is determined so as to make the left-hand
side of Eq.~(\ref{eq:tmat-prim-to-conv}) become the Niggli
cell.\cite{Niggli-1928, Gruber-1973, Krivy-1976,
  Grosse-Kunstleve-Niggli-2004, niggli-cell} For the Laue class $2/m$, the
$\mathbf{b}$ axis direction is set as $\mathbf{e}^\mathrm{pri}$ along
the two fold rotation axis by Eq.~(\ref{eq:axis1}). The $\mathbf{a}$ and
$\mathbf{c}$ axis directions are found to be perpendicular to
$\mathbf{e}^\mathrm{pri}$ by Eq.~(\ref{eq:axis2}).
Therefore $(\mathbf{e}_x, \mathbf{e}_y, \mathbf{e}_z) =
  (\mathbf{e}^\mathrm{ter},\mathbf{e}^\mathrm{pri},
  \mathbf{e}^\mathrm{sec})$.
For the Laue classes of $4/m$, $4/mmm$, $\bar{3}$, $\bar{3}m$, $6/m$, and
$6/mmm$, the $\mathbf{c}$ axis direction is set as
$\mathbf{e}^\mathrm{pri}$ along the four or three fold rotation axis by
Eq.~(\ref{eq:axis1}). $\mathbf{e}^\mathrm{sec}$ is found to be
perpendicular to $\mathbf{e}^\mathrm{pri}$ by Eq.~(\ref{eq:axis2}), and
$\mathbf{e}^\mathrm{ter}=\boldsymbol{W}^\mathrm{pri}\mathbf{e}^\mathrm{sec}$.
Therefore $(\mathbf{e}_x, \mathbf{e}_y, \mathbf{e}_z) =
  (\mathbf{e}^\mathrm{sec}, \mathbf{e}^\mathrm{ter}, \mathbf{e}^\mathrm{pri})$.
Among possible sets of $(\mathbf{e}^\mathrm{sec},
  \mathbf{e}^\mathrm{ter}, \mathbf{e}^\mathrm{pri})$, one having the
smallest $|\det(\mathbf{M}')| \neq 0$ is selected to avoid wrongly-centred
$(\mathbf{a}_\mathrm{c},
  \mathbf{b}_\mathrm{c},\mathbf{c}_\mathrm{c})$.
For the Laue classes of $mmm$, $m\bar{3}$, and $m\bar{3}m$, three axis
directions along two or four-fold rotation axes are determined by
Eq.~(\ref{eq:axis1}).
When $\det(\mathbf{M}') < 0$, the secondary and ternary axis directions
are swapped to make the system of basis vectors right-handed.

\begin{table}
  \caption{\label{tab:correction-mat} Correction matrices $\mathbf{M}$.}
  \begin{center}
    \begin{tabular}{lcc}
      $2/m$, $A \rightarrow C$               &  & $\mathbf{M}_{2/m,A} =
      \begin{pmatrix}
          0 & 0       & 1 \\
          0 & \bar{1} & 0 \\
          1 & 0       & 0
        \end{pmatrix}$                                                                       \\ \\
      $2/m$, $B \rightarrow C$               &  & $\mathbf{M}_{2/m,B} =
      \begin{pmatrix}
          0 & 1 & 0 \\
          0 & 0 & 1 \\
          1 & 0 & 0
        \end{pmatrix}$                                                                        \\ \\
      $2/m$, $I \rightarrow C$               &  & $\mathbf{M}_{2/m,I} =
      \begin{pmatrix}
          1 & 0 & \bar{1} \\
          0 & 1 & 0       \\
          1 & 0 & 0
        \end{pmatrix}$                                                                       \\ \\
      $mmm$, $A \rightarrow C$               &  & $\mathbf{M}_{mmm,A} =
      \begin{pmatrix}
          0 & 0 & 1 \\
          1 & 0 & 0 \\
          0 & 1 & 0
        \end{pmatrix}$                                                                        \\ \\
      $mmm$, $B \rightarrow C$               &  & $\mathbf{M}_{mmm,B} = \mathbf{M}_{2/m,B}$ \\ \\
      \makecell[l]{Obverse hexagonal cell \\ $\rightarrow$ primitive rhombohedral cell} &  & $\mathbf{M}_\mathrm{obv} =
        \renewcommand*{\arraystretch}{1.3}
      \begin{pmatrix}
          \frac{2}{3} & \bar{\frac{1}{3}} & \bar{\frac{1}{3}} \\
          \frac{1}{3} & \frac{1}{3}       & \bar{\frac{2}{3}} \\
          \frac{1}{3} & \frac{1}{3}       & \frac{1}{3}       \\
        \end{pmatrix}$           \\ \\
      \makecell[l]{Reverse hexagonal cell \\ $\rightarrow$ primitive rhombohedral cell} &  & $\mathbf{M}_\mathrm{rev} = \left(
        \renewcommand*{\arraystretch}{1.3}
        \begin{matrix}
          \frac{1}{3} & \bar{\frac{2}{3}} & \frac{1}{3}       \\
          \frac{2}{3} & \bar{\frac{1}{3}} & \bar{\frac{1}{3}} \\
          \frac{1}{3} & \frac{1}{3}       & \frac{1}{3}       \\
        \end{matrix}
      \right)$                                                                              \\ \\
      Otherwise,                             &  & $\mathbf{M} =
        \renewcommand*{\arraystretch}{1.3}
      \begin{pmatrix}
          1 & 0 & 0 \\
          0 & 1 & 0 \\
          0 & 0 & 1
        \end{pmatrix}$                                                                        \\ \\
    \end{tabular}
  \end{center}
\end{table}

For convenience in the following steps, the basis vectors are further
transformed to have a specific centering type by multiplying a correction matrix
$\mathbf{M}$ with $\mathbf{M}'$ for the Laue classes of $2/m$, $mmm$, and the
rhombohedral system. Otherwise, $\mathbf{M}$ is considered as an identity matrix.
For those, the correction matrices are listed in Table~\ref{tab:correction-mat}.
The current centering type is easily identified from $\mathbf{M}'$ using
Table.~\ref{tab:centering}.

\begin{table}
  \caption{\label{tab:centering} Conditions to determine the centering
    types.}
  \begin{center}
    \begin{tabular}{lcccc}
      Centering type      &  & $\det(\mathbf{M}')$ &  & Row vectors of
      $\mathbf{M}'$                                                                                                           \\
      \hline
      $P$                &  & 1                   &  & -                                                                      \\
      $A$                &  & 2                   &  & ${}^\exists i, \sum_j\left|M'_{ij}\right|=1$, $\left|M'_{i1}\right|=1$ \\
      $B$                &  & 2                   &  & ${}^\exists i, \sum_j\left|M'_{ij}\right|=1$, $\left|M'_{i2}\right|=1$ \\
      $C$                &  & 2                   &  & ${}^\exists i, \sum_j\left|M'_{ij}\right|=1$, $\left|M'_{i3}\right|=1$ \\
      $I$ (body)         &  & 2                   &  & ${}^\forall i, \sum_j \left|M'_{ij}\right|=2$                          \\
      $R$ (rhombohedral) &  & 3                   &  & -                                                                      \\
      $F$ (face)         &  & 4                   &  & -                                                                      \\
    \end{tabular}
  \end{center}
\end{table}

For the Laue class $2/m$, the basis vectors with the $I$, $A$, and $B$
centering types are transformed to those with the $C$ centering type. For
the Laue class $mmm$, those with the $A$, and $B$ centering types are
transformed to those with the $C$ centering type. For the rhombohedral
system, a rhombohedrally-centred hexagonal cell is obtained by
$\mathbf{M}'$ in either the obverse or reverse setting. This is
transformed to the primitive rhombohedral cell by
$\mathbf{M}_\mathrm{obv}$ if it is the obverse setting or by
$\mathbf{M}_\mathrm{rev}$ if it is the reverse setting. Only one of
$\mathbf{M}'\mathbf{M}_\mathrm{obv}$ or
$\mathbf{M}'\mathbf{M}_\mathrm{rev}$ has to be an integer matrix, which
is chosen as the transformation matrix to a primitive rhombohedral cell.

\subsection{Step (j): Identification of Hall symbol}

The space group operations obtained in the second stage are compared with the
datasets generated by decoding the Hall symbols. The Hall symbols are the
explicit-origin space group notation proposed by Hall in
Ref.~\onlinecite{Hall-1981}. The method to decode the Hall symbols is also
found in {\it International Tables for Crystallography Volume B}.\cite{ITB} The
530 sets of matrix representations are pre-decoded and stored in the
$\texttt{spglib}$ source code.

To perform the comparison, the set of the space group operations has to be
represented in the same coordinate system as that in the datasets. Using the
transformation matrix $\mathbf{M}'\mathbf{M}$ obtained in the step (i) as
described in Eq.~(\ref{eq:transform-operation}), the space group operations are
transformed into those corresponding to one specific conventional unit cell
setting.

To match those in different unique axes, settings, or cell choices described by
the Hall symbols, an additional change-of-basis matrix $\mathbf{Q}'$ is
employed. In the \texttt{spglib} code, the matrix $\mathbf{Q}'$ can be selected
such that after the transformation, it favors the conditions
$|\mathbf{a}_\mathrm{c}| \leq |\mathbf{b}_\mathrm{c}|$, $|\mathbf{a}_\mathrm{c}|
\leq |\mathbf{c}_\mathrm{c}|$, or $|\mathbf{b}_\mathrm{c}| \leq
|\mathbf{c}_\mathrm{c}|$, and being similar to the input unit cell in
orientation. This preference is permissible when the choice of change-of-basis
matrix is not constrained by the Hall symbol. Thus, the change-of-basis matrix
$\mathbf{M}'\mathbf{M}$ is updated by $\mathbf{Q}'$ to
$\mathbf{M}'\mathbf{M}\mathbf{Q}'$. For the space group type $Pa\bar{3}$, two
change-of-basis matrices, $\mathbf{M}'\mathbf{M}$ and ${\mathbf{M}'\mathbf{M}
\mathbf{Q}'}_{Pa\bar{3}}$, are examined to match,\cite{Grosse-Kunstleve-I-1999}
where $\mathbf{Q}_{Pa\bar{3}}'$ used in the \texttt{spglib} code is given by
\begin{equation}
  \mathbf{Q}'_{Pa\bar{3}} =
  \begin{pmatrix}
    0 & 0       & 1 \\
    0 & \bar{1} & 0 \\
    1 & 0       & 0
  \end{pmatrix}.
\end{equation}

\begin{table}
  \caption{\label{tab:trans-c2p} Transformation matrices used in the Hall
    symbol matching. These matrices transform conventional unit cell
    settings to respective primitive cell settings. The subscripts $X$ of
    the matrices $\mathbf{P}_X$ indicate the centering types: $A$, $B$, $C$
    for the base centering types, $I$ and $F$ for the body and face centering
    types, respectively, and $R$ for the (obverse) rhombohedral centering
    type. }
  \begin{center}
    \begin{tabular}{lll}
      $\mathbf{P}_A =
        \renewcommand*{\arraystretch}{1.3}
        \begin{pmatrix}
          1 & 0           & 0                 \\
          0 & \frac{1}{2} & \bar{\frac{1}{2}} \\
          0 & \frac{1}{2} & {\frac{1}{2}}     \\
        \end{pmatrix}
      $, &
      $\mathbf{P}_B =
        \renewcommand*{\arraystretch}{1.3}
        \begin{pmatrix}
          \frac{1}{2} & 0 & \bar{\frac{1}{2}} \\
          0           & 1 & 0                 \\
          \frac{1}{2} & 0 & {\frac{1}{2}}     \\
        \end{pmatrix}
      $, &
      $\mathbf{P}_C =
        \renewcommand*{\arraystretch}{1.3}
        \begin{pmatrix}
          \frac{1}{2}       & \frac{1}{2} & 0 \\
          \bar{\frac{1}{2}} & \frac{1}{2} & 0 \\
          0                 & 0           & 1 \\
        \end{pmatrix}
      $,   \\ \\
      $\mathbf{P}_I =
        \renewcommand*{\arraystretch}{1.3}
        \begin{pmatrix}
          \bar{\frac{1}{2}} & \frac{1}{2}       & \frac{1}{2}       \\
          {\frac{1}{2}}     & \bar{\frac{1}{2}} & \frac{1}{2}       \\
          {\frac{1}{2}}     & \frac{1}{2}       & \bar{\frac{1}{2}} \\
        \end{pmatrix}
      $, &
      $\mathbf{P}_F =
        \renewcommand*{\arraystretch}{1.3}
        \begin{pmatrix}
          0             & \frac{1}{2} & \frac{1}{2} \\
          {\frac{1}{2}} & 0           & \frac{1}{2} \\
          {\frac{1}{2}} & \frac{1}{2} & 0           \\
        \end{pmatrix}
      $, &
      $\mathbf{P}_R =
        \renewcommand*{\arraystretch}{1.3}
        \begin{pmatrix}
          \frac{2}{3} & \bar{\frac{1}{3}} & \bar{\frac{1}{3}} \\
          \frac{1}{3} & \frac{1}{3}       & \bar{\frac{2}{3}} \\
          \frac{1}{3} & \frac{1}{3}       & \frac{1}{3}       \\
        \end{pmatrix}
      $.   \\ \\
    \end{tabular}
  \end{center}
\end{table}

The matrix representations of the space group operations of
$\{(\boldsymbol{W}_\text{p},\boldsymbol{w}_\text{p})\}$ in the second stage are
given for the primitive cell. Each
$(\boldsymbol{W}_\text{p},\boldsymbol{w}_\text{p})$ is transformed to that in
the conventional unit cell, $(\boldsymbol{W}_\text{c},\boldsymbol{w}_\text{c})$,
by $\mathbf{M'MQ'}$ as described in Eq.~(\ref{eq:transform-operation}). After
this transformation, the rotation matrices of $\{\boldsymbol{W}_\text{c}\}$ are
directly comparable with those in the datasets.

To compare the translation parts to those in the datasets, their origins have to
be aligned. An origin shift is determined using generators of the space group
operations represented in the primitive cell setting. For the system having any
centering, $(\boldsymbol{W}_\text{c},\boldsymbol{w}_\text{c})$ is transformed to
$(\boldsymbol{W}_X,\boldsymbol{w}_X)$ by using the transformation matrix
$\mathbf{P}_X$ as presented in Table~\ref{tab:trans-c2p}.

When the space group operation is represented by
$(\boldsymbol{W}_X,\boldsymbol{w}_X)$ with reference to the origin $\mathbf{O}$
and by $(\boldsymbol{W}_X,\boldsymbol{w}^{\text{d}}_X)$ with respect to
$\mathbf{O}^{\text{d}}$, and both matrix representations describe the same
operation, they are interconnected by
\begin{equation}
  \label{eq:origin-shift-relation}
  (\boldsymbol{W}_X-\boldsymbol{I})
  \mathbf{p}_\text{p}=\boldsymbol{w}_X - \boldsymbol{w}_X^{\text{d}},
\end{equation}
where $\mathbf{p}_\text{p}$ is the origin shift from $\mathbf{O}^{\text{d}}$ to
$\mathbf{O}$, i.e., $\mathbf{O} = \mathbf{O}^{\text{d}} +
\mathbf{p}_\text{p}$.

Consider $(\boldsymbol{W}^\text{d}_X,\boldsymbol{w}^\text{d}_X)$ as the
reference provided from the dataset, and $(\boldsymbol{W}_X,\boldsymbol{w}_{X})$
as derived using this symmetry-finding algorithm, where
$\boldsymbol{W}^\text{d}_X=\boldsymbol{W}_X$. To determine
$\mathbf{p}_\text{p}$, at most three matrix representations of generators
are required. For example, using these three generators, we can solve the
equation below:
\begin{equation}
  \label{eq:origin-equation}
  % \mathbf{N}  \mathbf{c}_\text{p} =
  \left(
  \begin{matrix}
      \boldsymbol{W}_{X,\text{1}} - \boldsymbol{I} \\
      \boldsymbol{W}_{X,\text{2}} - \boldsymbol{I} \\
      \boldsymbol{W}_{X,\text{3}} - \boldsymbol{I} \\
    \end{matrix}
  \right)  \mathbf{p}_\text{p} =
  \left(
  \begin{matrix}
      \boldsymbol{w}_{X,\text{1}} - \boldsymbol{w}_{X,\text{1}}^{\text{d}} \\
      \boldsymbol{w}_{X,\text{2}} - \boldsymbol{w}_{X,\text{2}}^{\text{d}} \\
      \boldsymbol{w}_{X,\text{3}} - \boldsymbol{w}_{X,\text{3}}^{\text{d}} \\
    \end{matrix}
  \right) = \Delta \boldsymbol{w}_\text{p}\;(\bmod\;\mathbb{Z}).
\end{equation}
This is rewritten as
\begin{equation}
  \mathbf{N}\mathbf{p}_\text{p} =
  \Delta\boldsymbol{w}_\text{p}\;(\bmod\;\mathbb{Z}),
\end{equation}
where
\begin{equation}
  \mathbf{N} =
  \left(
  \begin{matrix}
      \boldsymbol{W}_{X,\text{1}} - \boldsymbol{I} \\
      \boldsymbol{W}_{X,\text{2}} - \boldsymbol{I} \\
      \boldsymbol{W}_{X,\text{3}} - \boldsymbol{I}
    \end{matrix}
  \right).
\end{equation}
A set of solutions is obtained by applying the Smith normal form $\mathbf{S}$
given by
\begin{equation}
  \label{eq:snf}
  \mathbf{S} = \mathbf{U} \mathbf{N} \mathbf{V},
\end{equation}
where $\mathbf{U}$ and $\mathbf{V}$ are the unimodular matrices. In the case
with the three generators, $\mathbf{N}$ is a $9\times 3$ matrix and its Smith
normal form $\mathbf{S}$ becomes
\begin{equation}
  \mathbf{S} =
  \left(
  \begin{matrix}
      {\lambda_1} & 0           & 0           \\
      0           & {\lambda_2} & 0           \\
      0           & 0           & {\lambda_3} \\
      0           & 0           & 0           \\
      0           & 0           & 0           \\
      0           & 0           & 0           \\
      0           & 0           & 0           \\
      0           & 0           & 0           \\
      0           & 0           & 0           \\
    \end{matrix}
  \right).
\end{equation}
The $3\times 9$ matrix $\mathbf{T}$ of the inverse
diagonal elements of $\mathbf{S}$ is made as
\begin{equation}
  \mathbf{T} =
  \left(
  \begin{matrix}
      \frac{1}{\lambda_1} & 0                   & 0                   & 0 & 0 & 0 & 0 & 0 & 0 \\
      0                   & \frac{1}{\lambda_2} & 0                   & 0 & 0 & 0 & 0 & 0 & 0 \\
      0                   & 0                   & \frac{1}{\lambda_3} & 0 & 0 & 0 & 0 & 0 & 0 \\
    \end{matrix}
  \right).
\end{equation}
When $\lambda_n=0$, the corresponding elements of $\mathbf{T}$ are set to 0.
Since $\mathbf{S} \mathbf{V}^{-1} \mathbf{p}_\text{p} = \mathbf{U}
\Delta\boldsymbol{w}_\text{p}$, one of the solutions is given
by
\begin{equation}
  \label{eq:VTU}
  \mathbf{p}_\text{p} = \mathbf{V} \mathbf{T} \mathbf{U}
  \Delta\boldsymbol{w}_\text{p}.
\end{equation}
A Python script was written to compute the matrices $\mathbf{V} \mathbf{T}
\mathbf{U}$  for crystal systems and their centering types and axis settings
using the \texttt{SageMath} code.\cite{SageMath} The precomputed $\mathbf{V}
\mathbf{T} \mathbf{U}$ matrices are stored in the \texttt{spglib} source code
together with the corresponding sets of the rotation parts of the generators in
the primitive cell setting.

With the origin shift $\mathbf{p}_\text{p}$ obtained in Eq.~(\ref{eq:VTU}), the
translation parts of the space group operations are compared with the datasets
by Eq.~(\ref{eq:origin-shift-relation}). Finally, the Hall symbol is idenfified
by verifying that $\{(\boldsymbol{W}_X,\boldsymbol{w}_X)\}$ is mapped to
$\{(\boldsymbol{W}^\text{d}_X,\boldsymbol{w}_X^\text{d})\}$.

For subsequent use, the basis vectors of the primitive cell as
obtained in Eq.~(\ref{eq:primitive-basis-vectors}) are transformed to those of
the conventional unit cell by
\begin{equation}
  \label{eq:prim-to-conv-basis-vectors}
  (\mathbf{a}_\text{c}, \mathbf{b}_\text{c}, \mathbf{c}_\text{c}) =
  (\mathbf{a}_\text{p}, \mathbf{b}_\text{p},
  \mathbf{c}_\text{p}) \mathbf{M}'\mathbf{M}\mathbf{Q}'.
\end{equation}
Furthermore, the origin shift in the coordinate system of the conventional unit
cell is given by
\begin{equation}
  \label{eq:conv-origin-shift}
  \mathbf{p}_\text{c} = \mathbf{P}_X \mathbf{p}_\text{p}.
\end{equation}

\subsection{Step (k): Failure of identification of space group type}

When the Hall symbol matching at step (j) fails, the tolerance value is
shortened and the current set of the space group operations is re-examined in
step (l). Following this, the sequence from steps (i) to (l) is reiterated. This
process continues until a Hall symbol is successfully identified. In cases where
this loop is executed numerous times without success, the entire procedure is
restarted from step (a), employing a tolerance value shortened from the one
utilized in the last attempt at step (a).

\subsection{Step (l): Thinning out space group operations}
The space group operations in
$\{(\boldsymbol{W}_{\text{p}},\boldsymbol{w}_{\text{p}})\}$ are re-evaluated in
a way similar to step (g) with the shortened tolerance value.  As a result,
certain space group operations in
$\{(\boldsymbol{W}_{\text{p}},\boldsymbol{w}_{\text{p}})\}$ may be excluded.
Since $\boldsymbol{w}_{\text{p}}$ is given, this re-examination requires
significantly less computational demand compared to the full execution of step
(g). Following this, the procedure revisits step (i) with the refined set of
space group operations.

\section{Finalization}
\label{sec:finalization}

In the fourth stage, the results from the previous stages are organized to
enhance their usefulness and intuitiveness. The matrix representations of the
space group operations for the input unit cell are reconstructed using the
transformation matrix, origin shift, and the Hall symbol dataset to minimize
distortions in the translational parts. Additionally, Wyckoff positions are
determined and a distortion-free crystal structure of the conventional unit
cell, which is derived from the input unit cell, is suggested.

\subsection{Step (m): Removal of distortion from basis vectors}

The tolerance introduced in the symmetry identification process may lead to some
distortion in $(\mathbf{a}_\text{c}, \mathbf{b}_\text{c}, \mathbf{c}_\text{c})$
compared to their ideal values for the determined lattice system. This
distortion is rectified by averaging the lengths of symmetrically equivalent
basis vectors to align with the expected values for the lattice system. The
specific lattice system conditions required for this adjustment are detailed in
Appendix~\ref{sec:symmetrization-of-basis-vectors}, where the basis
vectors $(\mathbf{a}_\text{s}, \mathbf{b}_\text{s}, \mathbf{c}_\text{s})$ are
adjusted for an intuitive alignment with the Cartesian axes. In this
rectification process, the basis vectors $(\mathbf{a}_\text{c},
\mathbf{b}_\text{c}, \mathbf{c}_\text{c})$ are subject to a rotation in
Cartesian coordinates, approximately represented by the equation:
\begin{align}
(\mathbf{a}_\text{s}, \mathbf{b}_\text{s}, \mathbf{c}_\text{s}) \approx
(\boldsymbol{R}\mathbf{a}_\text{c}, \boldsymbol{R}\mathbf{b}_\text{c},
\boldsymbol{R}\mathbf{c}_\text{c}),
\end{align}
where $(\mathbf{a}_\text{s}, \mathbf{b}_\text{s}, \mathbf{c}_\text{s})$ represent
the idealized basis vectors of the conventional unit cell.

\subsection{Step (n): Reconstruction of matrix representations of space
  group operations of input unit cell}
\label{sec:spgroup-unitcell}

The set of matrix representations of the space group operations
corresponding to the Hall symbol, $\{(\boldsymbol{W}^\text{d},
\boldsymbol{w}^\text{d})\}$, is obtained from the dataset. This is transformed
to that of the input unit cell $\{(\boldsymbol{W}_\text{i},
\boldsymbol{w}_\text{i})\}$ using the following procedure.

The basis vectors $(\mathbf{a}_\text{c}, \mathbf{b}_\text{c},
\mathbf{c}_\text{c})$, as specified by
Eq.~(\ref{eq:prim-to-conv-basis-vectors}), follow the symmetry of
$\{(\boldsymbol{W}^\text{d}, \boldsymbol{w}^\text{d})\}$. However, the basis
vectors resulting from applying the rotations $\boldsymbol{W}^\text{d*}$
are also valid:
\begin{align}
  (\mathbf{a}_\text{c}', \mathbf{b}_\text{c}', \mathbf{c}_\text{c}')  =
  (\mathbf{a}_\text{c}, \mathbf{b}_\text{c},
  \mathbf{c}_\text{c})\boldsymbol{W}^\text{d*}.
\end{align}
Consequently, as a potentially more suitable choice of basis vectors,
$(\mathbf{a}_\text{c}, \mathbf{b}_\text{c}, \mathbf{c}_\text{c})$ may be replaced
by $(\mathbf{a}_\text{c}', \mathbf{b}_\text{c}', \mathbf{c}_\text{c}')$. This
replacement aims to minimize the following expression:
\begin{align}
  |\mathbf{a}_\text{s} - \mathbf{a}_\text{c}'|^2 +
  |\mathbf{b}_\text{s} - \mathbf{b}_\text{c}'|^2 +
  |\mathbf{c}_\text{s} - \mathbf{c}_\text{c}'|^2.
\end{align}
As this involves a transformation of the coordinate system,
$\mathbf{p}_\text{c}$ as given by Eq.~(\ref{eq:conv-origin-shift}) is also
transformed according to:
\begin{align}
\mathbf{p}_\text{c}' = {\boldsymbol{W}^\text{d*}}^{-1}
(\mathbf{p}_\text{c} -\boldsymbol{w}^\text{d*}).
\end{align}
Details of this coordinate transformation are provided in
Appendix~\ref{sec:althernative-origin-shift}. For conventional unit cells having
any centering, $(\boldsymbol{W}^\text{d*}, \boldsymbol{w}^\text{d*})$ that gives
the shortest $\mathbf{p}_\text{c}'$ is employed. With the use of
$\mathbf{p}_\text{c}'$, the translation parts of $\{(\boldsymbol{W}^\text{d},
\boldsymbol{w}^\text{d})\}$ are redefined as:
\begin{equation}
\label{eq:origin-equation-conv}
{\boldsymbol{w}^\text{d}}' = (\boldsymbol{W}^\text{d} - \boldsymbol{I})
\mathbf{p}_\text{c}' + {\boldsymbol{w}^\text{d}}.
\end{equation}

Every $(\boldsymbol{W}^\text{d}, {\boldsymbol{w}^\text{d}}')$ is transformed to
that of the primitive cell $(\boldsymbol{W}^\text{d}_\text{p},
{\boldsymbol{w}^{\text{d}}_\text{p}}')$ as the transformation detailed in
Appendix~\ref{sec:trans-mat}. The transformation matrix
$\mathbf{Q}_{\text{c}\rightarrow\text{p}}$ is defined by the equation:
\begin{equation}
  \label{eq:p-to-c-transformation}
  (\mathbf{a}_\text{p}, \mathbf{b}_\text{p}, \mathbf{c}_\text{p}) =
  (\mathbf{a}_\text{c}', \mathbf{b}_\text{c}',
  \mathbf{c}_\text{c}') \mathbf{Q}_{\text{c}\rightarrow\text{p}}.
\end{equation}
Note that $\mathbf{Q}_{\text{c}\rightarrow\text{p}}^{-1}$ is an integer matrix.
To ensure that $\{(\boldsymbol{W}^\text{d}_\text{p},
{\boldsymbol{w}^{\text{d}}_\text{p}}')\}$ represents the set of coset
representatives, only one space group operation is included in
$\{(\boldsymbol{W}^\text{d}_\text{p}, {\boldsymbol{w}^{\text{d}}_\text{p}}')\}$ for
each rotation part $\boldsymbol{W}^\text{d}_\text{p}$.

Then, $\{(\boldsymbol{W}^\text{d}_\text{p},
{\boldsymbol{w}^{\text{d}}_\text{p}}')\}$ is transformed to that of the input
unit cell $\{(\boldsymbol{W}_\text{i,p} \boldsymbol{w}_\text{i,p})\}$. The
transformation matrix $\mathbf{Q}_{\text{p}\rightarrow\text{i}}$ is defined by
the equation:
\begin{equation}
  \label{eq:standardize-basis-vectors}
  (\mathbf{a}_\text{i}, \mathbf{b}_\text{i},
  \mathbf{c}_\text{i})  \approx
  (\mathbf{a}_\text{p}, \mathbf{b}_\text{p}, \mathbf{c}_\text{p})
  \mathbf{Q}_{\text{p}\rightarrow\text{i}},
\end{equation}
where the elements of the matrix $\mathbf{Q}_{\text{p}\rightarrow\text{i}}$ are
rounded to their nearest integer values, thereby constituting an integer matrix.
Thus obtained $\boldsymbol{W}_\text{i,p}$ may not be an integer matrix if the
order of the lattice point group of $(\mathbf{a}_\text{i}, \mathbf{b}_\text{i},
\mathbf{c}_\text{i})$ is smaller than that of $(\mathbf{a}_\text{p},
\mathbf{b}_\text{p}, \mathbf{c}_\text{p})$. Those space group operations with
non-integer matrices of $\boldsymbol{W}_\text{i,p}$ are excluded from
$\{(\boldsymbol{W}_\text{i,p}, \boldsymbol{w}_\text{i,p})\}$.

When the input unit cell is not a primitive cell, $\{(\boldsymbol{W}_\text{i,p},
\boldsymbol{w}_\text{i,p})\}$ is extended by a set of lattice point vectors in
the input unit cell, $\{\boldsymbol{t}_j\}$,
\begin{equation}
  \{(\boldsymbol{W}_\text{i}, \boldsymbol{w}_\text{i})\} =
  \bigcup_{j} (\boldsymbol{I}, \boldsymbol{t}_j)
  \{(\boldsymbol{W}_\text{i,p},
  \boldsymbol{w}_\text{i,p})\}.
\end{equation}
The lattice point vectors of $\{\boldsymbol{t}_i\}$
are easily obtained from
$\mathbf{Q}_{\text{p}\rightarrow\text{i}}$.\cite{phonopy-phono3py-JPCM}

\subsection{Step (o): Removal of distortion from point coordinates
  and determination of Wyckoff positions}

Point coordinates in the primitive cell $\boldsymbol{x}_\text{p}$ are
transformed to those in the conventional unit cell $\boldsymbol{x}_\text{c}$ by
\begin{equation}
  \boldsymbol{x}_\text{c} = \mathbf{Q}_{\text{c}\rightarrow\text{p}}
  \boldsymbol{x}_\text{p} + \mathbf{p}_\text{c}',
\end{equation}
Applying $\{(\boldsymbol{W}^\text{d}, \boldsymbol{w}^\text{d})\}$ to
$\{\boldsymbol{x}_\text{c}\}$, symmetrically independent points and sets of
symmetrically equivalent points are obtained.

Site symmetry operations of $\{(\boldsymbol{W}_{\boldsymbol{x}, i},
  \boldsymbol{w}_{\boldsymbol{x}, i})\}$ at $\boldsymbol{x}_\text{c}$ are the space
group operations that leave coordinates of a point $\boldsymbol{x}_\text{c}$
unchanged, i.e.,
\begin{equation}
  \label{eq:site-symmetry}
  (\boldsymbol{W}_{\boldsymbol{x}, i}, \boldsymbol{w}_{\boldsymbol{x}, i})
  \boldsymbol{x}_\text{c} = \boldsymbol{x}_\text{c}.
\end{equation}
$\{(\boldsymbol{W}_{\boldsymbol{x}, i},
\boldsymbol{w}_{\boldsymbol{x}, i})\}$ is expected to form the site symmetry group
$\mathbb{S}_{\boldsymbol{x}}$ of the finite order
$|\mathbb{S}_{\boldsymbol{x}}|$. Using
$\mathbb{S}_{\boldsymbol{x}}$, the special position
operator~\cite{Grosse-Kunstleve-II-2002}
$(\boldsymbol{W}_{\boldsymbol{x}}^\text{sp},
\boldsymbol{w}_{\boldsymbol{x}}^\text{sp})$ is defined as
\begin{equation}
  \label{eq:site-symmetry-ave}
  (\boldsymbol{W}_{\boldsymbol{x}}^\text{sp},
  \boldsymbol{w}_{\boldsymbol{x}}^\text{sp}) =
  \frac{1}{|\mathbb{S}_{\boldsymbol{x}}|}
  \sum_{i=1}^{|\mathbb{S}_{\boldsymbol{x}}|}(\boldsymbol{W}_{\boldsymbol{x},i},
  \boldsymbol{w}_{\boldsymbol{x},i}).
\end{equation}
Point coordinates $\boldsymbol{x}_\text{c}$ can be slightly dislocated from the
exact location. By Eq.~(\ref{eq:site-symmetry-ave}), the exact location
$\boldsymbol{x}_\mathbb{S}$ of $\boldsymbol{x}_\text{c}$ is obtained by
\begin{equation}
  \label{eq:proj-special-pos-operator}
  \boldsymbol{x}_\mathbb{S} = (\boldsymbol{W}_{\boldsymbol{x}}^\text{sp},
  \boldsymbol{w}_{\boldsymbol{x}}^\text{sp}) \boldsymbol{x}_\text{c}.
\end{equation}

In the \texttt{spglib} implementation, $\{(\boldsymbol{W}_{\boldsymbol{x}, i},
  \boldsymbol{w}_{\boldsymbol{x}, i})\}$ is obtained from
  $\{(\boldsymbol{W}^\text{d}, \boldsymbol{w}^\text{d})\}$. Since
  $\{(\boldsymbol{W}^\text{d}, \boldsymbol{w}^\text{d})\}$ is the coset
  representatives of the lattice translation group of the conventional unit
  cell, i.e., it is a finite set not like a space group,
  Eq.~(\ref{eq:site-symmetry}) is examined using the
  condition~(\ref{eq:red-Euclidean-dist}) as
\begin{align}
  \label{eq:site-distortion}
  \Delta \boldsymbol{x}_i =
  (\boldsymbol{W}_{\boldsymbol{x},i}, \boldsymbol{w}_{\boldsymbol{x},i})
  \boldsymbol{x}_\text{c} - \boldsymbol{x}_\text{c}.
\end{align}
Using Eq.~(\ref{eq:site-distortion}),
Eqs.~(\ref{eq:site-symmetry-ave}) and
 (\ref{eq:proj-special-pos-operator}) are rewritten as
\begin{equation}
  \boldsymbol{x}_\mathbb{S} = \frac{1}{|\mathbb{S}_{\boldsymbol{x}}|}
  \sum_i^{|\mathbb{S}_{\boldsymbol{x}}|}
  (\Delta \boldsymbol{x}_i - \lfloor\Delta\boldsymbol{x}_i\rceil) +
  \boldsymbol{x}_\mathrm{c}.
\end{equation}

The number of the symmetrically equivalent points of the point
$\boldsymbol{x}_\mathbb{S}$ in the conventional
unit cell is called multiplicity $M_{\boldsymbol{x}}$. Note that
$M_{\boldsymbol{x}}$ is defined with respect to the conventional unit cell but
not the primitive cell as following the convention of the {\it International
Tables for Crystallography Volume A}.\cite{ITA} These have to satisfy the
following relationship:
\begin{equation}
  \label{eq:multiplicity-orbit}
  |\mathbb{S}_{\boldsymbol{x}} |M_{\boldsymbol{x}} = |\mathbb{S}/\mathbb{T}|
  \det(\mathbf{P}_X^{-1}),
\end{equation}
where $|\mathbb{S}/\mathbb{T}|$ denotes the order of the factor group. This is
same as the cardinality of the coset representatives obtained for the primitive
cell. Obviously $\det(\mathbf{P}_X^{-1})$ is equivalent to the number of the
lattice points in the conventional unit cell. Finally,
$\boldsymbol{x}_\mathbb{S}$ of the symmetrically independent points are
obtained, and these points are then expanded to their symmetrically equivalent
points.

The Wyckoff letter of $\boldsymbol{x}_\mathbb{S}$ is determined using {\it
  Coordinates} in the Wyckoff position dataset. {\it Coordinates} are listed in
  the {\it International Tables for Crystallography Volume A}.\cite{ITA} The
  dataset in the \texttt{spglib} code was provided by Y.
  Seto~\cite{seto-database} for all the Hall symbols. The first entries of {\it
  Coordinates} of Wyckoff positions for each Hall symbol is necessary to match
  $\boldsymbol{x}_\mathbb{S}$ with a Wyckoff letter. All those first entries of
  {\it Coordinates} are stored in the \texttt{spglib} source code in a matrix
  format. For example, the first entry of the Wyckoff letter $f$ of $P42_12$ (No.
  90) is $(x, x, \frac{1}{2})$, which is represented by
\begin{equation}
  \begin{pmatrix}
    1 & 0 & 0 \\
    1 & 0 & 0 \\
    0 & 0 & 0
  \end{pmatrix}
  \begin{pmatrix}
    x \\
    y \\
    z
  \end{pmatrix} +
  \begin{pmatrix}
    0 \\
    0 \\
    \frac{1}{2}
  \end{pmatrix}. \nonumber
\end{equation}
The $3\times 3$ and $3\times 1$ matrices are encoded and stored in the
\texttt{spglib} soruce code. Matching with the dataset is performed by examining
$(x, x, \frac{1}{2})\boldsymbol{x}_\mathbb{S} = \boldsymbol{x}_\mathbb{S}
\;(\bmod\;\mathbb{Z})$. The multiplicity $M_{\boldsymbol{x}}$ is also stored in
the dataset. With this $M_{\boldsymbol{x}}$, Eq.~(\ref{eq:multiplicity-orbit})
is verified. If this fails, the procedure restarts from step (a) with the
tolerance value shortened from that used last time at step (a).

\section{Summary}

The \texttt{spglib} code is designed for the identification and symmetrization
of crystal structures, that are provided as basis vectors, point coordinates,
and atomic types, tolerating slight distortion. Utilizing established
crystallography knowledge and algorithms, it examines crystal symmetry and
validates symmetry operations searched numerically. During this process, an
input tolerance value is adjusted to align matrix representations of symmetry
operations with one of the space group types.

As the development of the \texttt{spglib} code has evolved, its source code has
become less readable due to the series of incremental improvements made over
time. This paper aims to clarify the implementation strategy of the current
version of the \texttt{spglib} code, particularly for those keen on
understanding the framework. Therefore, every detail is thoroughly described as
it is implemented.

The accumulation of technical debt has complicated code maintenance,
necessitating periodic major updates to support sustained scientific progress.
This will be achieved by selecting a suitable programming language for
each respective situation along with keeping the core of the code concise and
efficient.

\section*{ACKNOWLEDGMENTS}
We would like to thank Cristian Le and P. T. Jochym for their dedication to the
continuous integration system of the \texttt{spglib} code. We also would like to
thank Y. Seto for providing the space group type and Wyckoff position datasets.
This work was supported by JSPS KAKENHI Grant Number JP21K04632.

\appendix

\section{Notations and conventions}
\label{sec:notations}

In this section, the notations and conventions used in this paper are
summarized. Basically, we follow and respect the notations and
conventions of {\it International Tables for Crystallography Volume
    A}~\cite{ITA} and Refs.~\onlinecite{Grosse-Kunstleve-I-1999,
  Grosse-Kunstleve-II-2002}.

\subsection{Basis vectors $(\mathbf{a}, \mathbf{b}, \mathbf{c})$}

\label{sec:basis-vectors}

Basis vectors are represented by three column vectors:
\begin{align}
  \mathbf{a}= \begin{pmatrix}
                a_x \\
                a_y \\
                a_z \\
              \end{pmatrix},
  \mathbf{b}= \begin{pmatrix}
                b_x \\
                b_y \\
                b_z \\
              \end{pmatrix},
  \mathbf{c}= \begin{pmatrix}
                c_x \\
                c_y \\
                c_z \\
              \end{pmatrix},
\end{align}
in the Cartesian coordinates.

\subsection{Atomic point coordinates $\boldsymbol{x}$}

Coordinates of an atomic point $\boldsymbol{x}$ are represented
by three values relative to basis vectors as follows:
\begin{align}
  \boldsymbol{x}= \begin{pmatrix}
                    x_1 \\
                    x_2 \\
                    x_3 \\
                  \end{pmatrix}.
\end{align}
A position vector $\mathbf{x}$ in
the Cartesian coordinates are obtained by
\begin{align}
  \mathbf{x} = (\mathbf{a}, \mathbf{b}, \mathbf{c}) \boldsymbol{x}.
\end{align}

\subsection{Metric tensor $\boldsymbol{G}$ and point-group operation of
  lattice $\boldsymbol{W}$}
\label{sec:pg-lattice}
The metric tensor is defined by
\begin{equation}
  \label{eq:metric-tensor}
  \boldsymbol{G} = (\mathbf{a}, \mathbf{b}, \mathbf{c})^\top
  (\mathbf{a}, \mathbf{b}, \mathbf{c}).
\end{equation}
A rotation matrix $\boldsymbol{W}$ is applied to the basis
vectors such as,
\begin{equation}
  \label{eq:transformation}
  (\tilde{\mathbf{a}}, \tilde{\mathbf{b}}, \tilde{\mathbf{c}}) =
  (\mathbf{a}, \mathbf{b}, \mathbf{c}) \boldsymbol{W}.
\end{equation}
The metric tensor of $(\tilde{\mathbf{a}}, \tilde{\mathbf{b}},
  \tilde{\mathbf{c}})$ is given by
\begin{align}
  \label{eq:rotated-metric-tensor}
  \tilde{\boldsymbol{G}} & =
  (\tilde{\mathbf{a}}, \tilde{\mathbf{b}}, \tilde{\mathbf{c}})^\top
  (\tilde{\mathbf{a}}, \tilde{\mathbf{b}}, \tilde{\mathbf{c}}) \nonumber                              \\
                         & = \boldsymbol{W}^\top(\mathbf{a}, \mathbf{b}, \mathbf{c})^\top
  (\mathbf{a}, \mathbf{b}, \mathbf{c}) \boldsymbol{W} \nonumber                                       \\
                         & = \boldsymbol{W}^\top \boldsymbol{G} \boldsymbol{W}.
\end{align}
The lattice point group operations are obtained by searching matrices
$\{\boldsymbol{W}\}$ that satisfies $\boldsymbol{G}=\boldsymbol{W}^\top
  \boldsymbol{G} \boldsymbol{W}$, where $\boldsymbol{W}$ is the integer matrix
and the determinant is either 1 or $-1$.

\subsection{Space group operation $(\boldsymbol{W},\boldsymbol{w})$}

\label{sec:sp-operation}

A crystal structure is transformed by a space group operation in which
the coordinate system is at rest. Instead of rotating basis vectors as
given in Eq.~(\ref{eq:transformation}), a point in direct space,
$\boldsymbol{x}$, is transformed to a point $\tilde{\boldsymbol{x}}$ by
a rotation by
\begin{equation}
  \tilde{\boldsymbol{x}} = \boldsymbol{W}\boldsymbol{x}.
\end{equation}
A space group operation has a rotation part $\boldsymbol{W}$ and a
translation part $\boldsymbol{w}$. This is represented by the Seitz
symbol $(\boldsymbol{W},\boldsymbol{w})$ that transforms
$\boldsymbol{x}$ to $\tilde{\boldsymbol{x}}$ as
\begin{equation}
  \label{eq:spg-operation}
  \tilde{\boldsymbol{x}}=(\boldsymbol{W},\boldsymbol{w})
  \boldsymbol{x}= \boldsymbol{W}\boldsymbol{x}
  +\boldsymbol{w}.
\end{equation}
$\boldsymbol{W}$ and $\boldsymbol{w}$ are represented by a $3\times3$ integer
matrix and a $3\times 1$ column matrix, respectively. The point
$\tilde{\boldsymbol{x}}$ has to be equal to one of the points $\boldsymbol{x}$
in the primitive cell up to lattice translation, for
$(\boldsymbol{W},\boldsymbol{w})$ to be a space group operation.

\subsection{Transformation of coordinate system $(\boldsymbol{P},
    \boldsymbol{p})$}
\label{sec:transformation-spg}

The coordinate system of a crystal structure at rest is transformed by a pair
of a $3\times 3$ matrix $\boldsymbol{P}$ and a $3\times 1$ column matrix
$\boldsymbol{p}$, which is denoted by
$(\boldsymbol{P},\boldsymbol{p})$. The transformation matrix
$\boldsymbol{P}$ changes a choice of basis vectors as follows
\begin{align}
  \label{eq:transformation-of-basis}
  (\mathbf{a},\mathbf{b},\mathbf{c})
  = (\mathbf{a}_\mathrm{s},\mathbf{b}_\mathrm{s},\mathbf{c}_\mathrm{s})
  \boldsymbol{P},
\end{align}
where $(\mathbf{a},\mathbf{b},\mathbf{c} )$ and
$(\mathbf{a}_\mathrm{s},\mathbf{b}_\mathrm{s}, \mathbf{c}_\mathrm{s})$
are, e.g., the basis vectors of a primitive cell and those of the
conventional unit cell, respectively. The transformation matrix doesn't
rotate a crystal in the Cartesian coordinates, but just changes the
choices of the basis vectors. The origin shift $\boldsymbol{p}$ gives
the vector from the origin of an original coordinate system
$\boldsymbol{O}_\mathrm{s}$ to that of any other coordinate system
$\boldsymbol{O}$, which is written as
\begin{align}
  \boldsymbol{p} = \boldsymbol{O} - \boldsymbol{O}_\mathrm{s}.
\end{align}
The origin shift does not move the crystal in the Cartesian
coordinates, but just change the origin to measure the point coordinates.

The point coordinates in the original coordinate system
$\boldsymbol{x}_\mathrm{s}$ and those in the transformed coordinate
system $\boldsymbol{x}$ are related by
\begin{align}
  \label{eq:transformation-of-coordinates}
  \boldsymbol{x}_\mathrm{s} =
  (\boldsymbol{P},\boldsymbol{p})\boldsymbol{x} =
  \boldsymbol{P}\boldsymbol{x} +
  \boldsymbol{p},
\end{align}
where $\boldsymbol{p}$ is given with respect to the original basis
vectors. Equivalently,
\begin{align}
  \boldsymbol{x} = \boldsymbol{P}^{-1}\boldsymbol{x}_\mathrm{s} -
  \boldsymbol{P}^{-1}\boldsymbol{p}.
\end{align}

\begin{figure}[ht]
  \begin{center}
    \includegraphics[width=0.70\linewidth]{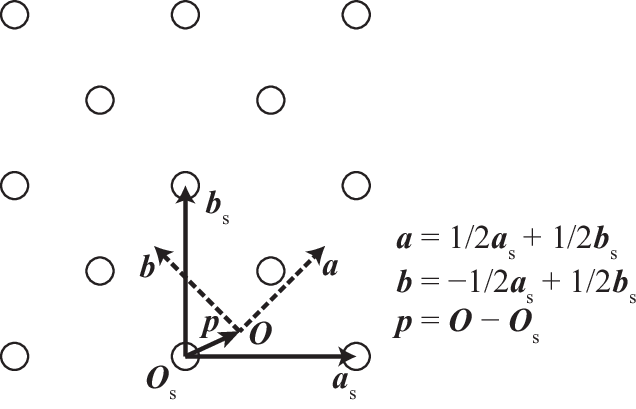}
    \caption{\label{fig:change-of-basis}
      Transformation of coordinate system.}
  \end{center}
\end{figure}

An illustration is presented in Fig.~\ref{fig:change-of-basis}. In this
case, the following $\boldsymbol{P}$ is applied:
\begin{align}
  \renewcommand*{\arraystretch}{1.4}
  \boldsymbol{P} = \begin{pmatrix}
                     \frac{1}{2}       & \frac{1}{2} & 0 \\
                     \frac{\bar{1}}{2} & \frac{1}{2} & 0 \\
                     0                 & 0           & 1
                   \end{pmatrix}.
\end{align}

\section{Crystallographic point group}
\label{sec:cryst-point-group}

The crystallographic point group type (crystal class) is uniquely
determined from matrix representations of the rotation parts of the
coset representatives.\cite{ITA} The rotation type is identified using
Table \ref{tab:rotations} with the determinant and trace of
$\boldsymbol{W}$. The crystal class is found comparing the list of
numbers of the rotation types with Table
\ref{tab:crystal-class}. Crystal system and Laue class are uniquely
assigned from the crystal class.

\begin{table}
  \caption{\label{tab:rotations} Look-up table to identify the types of
    the rotation parts of the space group operations from their matrix
    representations.}
  \begin{center}
    \begin{tabular}{lcrrrrrrrrrrrrrrrrrrr}
      Type of $\boldsymbol{W}$      &  & -6 &  & -4 &  & -3 &  & -2 &  & -1 &  & 1 &  & 2  &  & 3 &  & 4 &  & 6 \\
      \hline
      $\mathrm{tr}(\boldsymbol{W})$ &  & -2 &  & -1 &  & 0  &  & 1  &  & -3 &  & 3 &  & -1 &  & 0 &  & 1 &  & 2 \\
      $\det(\boldsymbol{W})$        &  & -1 &  & -1 &  & -1 &  & -1 &  & -1 &  & 1 &  & 1  &  & 1 &  & 1 &  & 1 \\
    \end{tabular}
  \end{center}
\end{table}

\begin{table}
  \caption{\label{tab:crystal-class} Look-up table of crystal class with
    the numbers of rotation types.}
  \begin{center}
    \begin{tabular}{lclclcrrrrrrrrrr}
      \hline
      Crystal                       &   & Crystal     &   & Laue                         &   & \multicolumn{10}{c}{Numbers of types of $\boldsymbol{W}$}                                         \\
      system                        &   & class       &   & class                        &   & -6                                                        & -4 & -3 & -2 & -1 & 1 & 2 & 3 & 4 & 6 \\
      \hline
      \multirow{2}{*}{Triclinic}    &   & $1$         &   & \multirow{2}{*}{$\bar{1}$}   &   &
      0                             & 0 & 0           & 0 & 0                            & 1 & 0                                                         & 0  & 0  & 0                           \\
                                    &   & $\bar{1}$   &   &                              &   &
      0                             & 0 & 0           & 0 & 1                            & 1 & 0                                                         & 0  & 0  & 0                           \\
      \hline
      \multirow{3}{*}{Monoclinic}   &   & $2$         &   & \multirow{3}{*}{$2/m$}       &   &
      0                             & 0 & 0           & 0 & 0                            & 1 & 1                                                         & 0  & 0  & 0                           \\
                                    &   & $m$         &   &                              &   &
      0                             & 0 & 0           & 1 & 0                            & 1 & 0                                                         & 0  & 0  & 0                           \\
                                    &   & $2/m$       &   &                              &   &
      0                             & 0 & 0           & 1 & 1                            & 1 & 1                                                         & 0  & 0  & 0                           \\
      \hline
      \multirow{3}{*}{Orthorhombic} &   & $222$       &   & \multirow{3}{*}{$mmm$}       &   &
      0                             & 0 & 0           & 0 & 0                            & 1 & 3                                                         & 0  & 0  & 0                           \\
                                    &   & $mm2$       &   &                              &   &
      0                             & 0 & 0           & 2 & 0                            & 1 & 1                                                         & 0  & 0  & 0                           \\
                                    &   & $mmm$       &   &                              &   &
      0                             & 0 & 0           & 3 & 1                            & 1 & 3                                                         & 0  & 0  & 0                           \\
      \hline
      \multirow{7}{*}{Tetragonal}   &   & $4$         &   & \multirow{3}{*}{$4/m$}       &   &
      0                             & 0 & 0           & 0 & 0                            & 1 & 1                                                         & 0  & 2  & 0                           \\
                                    &   & $\bar{4}$   &   &                              &   &
      0                             & 2 & 0           & 0 & 0                            & 1 & 1                                                         & 0  & 0  & 0                           \\
                                    &   & $4/m$       &   &                              &   &
      0                             & 2 & 0           & 1 & 1                            & 1 & 1                                                         & 0  & 2  & 0                           \\ \cline{2-16}
                                    &   & $422$       &   & \multirow{4}{*}{$4/mmm$}     &   &
      0                             & 0 & 0           & 0 & 0                            & 1 & 5                                                         & 0  & 2  & 0                           \\
                                    &   & $4mm$       &   &                              &   &
      0                             & 0 & 0           & 4 & 0                            & 1 & 1                                                         & 0  & 2  & 0                           \\
                                    &   & $\bar{4}2m$ &   &                              &   &
      0                             & 2 & 0           & 2 & 0                            & 1 & 3                                                         & 0  & 0  & 0                           \\
                                    &   & $4/mmm$     &   &                              &   &
      0                             & 2 & 0           & 5 & 1                            & 1 & 5                                                         & 0  & 2  & 0                           \\
      \hline
      \multirow{5}{*}{Trigonal}     &   & $3$         &   & \multirow{2}{*}{$\bar{3}$}   &   &
      0                             & 0 & 0           & 0 & 0                            & 1 & 0                                                         & 2  & 0  & 0                           \\
                                    &   & $\bar{3}$   &   &                              &   &
      0                             & 0 & 2           & 0 & 1                            & 1 & 0                                                         & 2  & 0  & 0                           \\ \cline{2-16}
                                    &   & $32$        &   & \multirow{3}{*}{$\bar{3}m$}  &   &
      0                             & 0 & 0           & 0 & 0                            & 1 & 3                                                         & 2  & 0  & 0                           \\
                                    &   & $3m$        &   &                              &   &
      0                             & 0 & 0           & 3 & 0                            & 1 & 0                                                         & 2  & 0  & 0                           \\
                                    &   & $\bar{3}m$  &   &                              &   &
      0                             & 0 & 2           & 3 & 1                            & 1 & 3                                                         & 2  & 0  & 0                           \\
      \hline
      \multirow{7}{*}{Hexagonal}    &   & $6$         &   & \multirow{3}{*}{$6/m$}       &   &
      0                             & 0 & 0           & 0 & 0                            & 1 & 1                                                         & 2  & 0  & 2                           \\
                                    &   & $\bar{6}$   &   &                              &   &
      2                             & 0 & 0           & 1 & 0                            & 1 & 0                                                         & 2  & 0  & 0                           \\
                                    &   & $6/m$       &   &                              &   &
      2                             & 0 & 2           & 1 & 1                            & 1 & 1                                                         & 2  & 0  & 2                           \\ \cline{2-16}
                                    &   & $622$       &   & \multirow{4}{*}{$6/mmm$}     &   &
      0                             & 0 & 0           & 0 & 0                            & 1 & 7                                                         & 2  & 0  & 2                           \\
                                    &   & $6mm$       &   &                              &   &
      0                             & 0 & 0           & 6 & 0                            & 1 & 1                                                         & 2  & 0  & 2                           \\
                                    &   & $\bar{6}2m$ &   &                              &   &
      2                             & 0 & 0           & 4 & 0                            & 1 & 3                                                         & 2  & 0  & 0                           \\
                                    &   & $6/mmm$     &   &                              &   &
      2                             & 0 & 2           & 7 & 1                            & 1 & 7                                                         & 2  & 0  & 2                           \\
      \hline
      \multirow{5}{*}{Cubic}        &   & $23$        &   & \multirow{2}{*}{$m\bar{3}$}  &   &
      0                             & 0 & 0           & 0 & 0                            & 1 & 3                                                         & 8  & 0  & 0                           \\
                                    &   & $m\bar{3}$  &   &                              &   &
      0                             & 0 & 8           & 3 & 1                            & 1 & 3                                                         & 8  & 0  & 0                           \\ \cline{2-16}
                                    &   & $432$       &   & \multirow{3}{*}{$m\bar{3}m$} &   &
      0                             & 0 & 0           & 0 & 0                            & 1 & 9                                                         & 8  & 6  & 0                           \\
                                    &   & $\bar{4}3m$ &   &                              &   &
      0                             & 6 & 0           & 6 & 0                            & 1 & 3                                                         & 8  & 0  & 0                           \\
                                    &   & $m\bar{3}m$ &   &                              &   &
      0                             & 6 & 8           & 9 & 1                            & 1 & 9                                                         & 8  & 6  & 0                           \\
      \hline
    \end{tabular}
  \end{center}
\end{table}

\section{Transformation of matrix representation of
  space group operation}
\label{sec:trans-mat}

When basis vectors $(\mathbf{a}_\mathrm{i},
  \mathbf{b}_\mathrm{i},\mathbf{c}_\mathrm{i})$ are transformed to another
basis vectors $(\mathbf{a}_\mathrm{f},
  \mathbf{b}_\mathrm{f},\mathbf{c}_\mathrm{f})$ by a change-of-basis
matrix $\mathbf{Q}$ as
\begin{equation}
  (\mathbf{a}_\mathrm{f},
  \mathbf{b}_\mathrm{f},\mathbf{c}_\mathrm{f}) = (\mathbf{a}_\mathrm{i},
  \mathbf{b}_\mathrm{i},\mathbf{c}_\mathrm{i}) \mathbf{Q},
\end{equation}
the matrix representation of the space group operation
$(\boldsymbol{W}_\mathrm{i},\boldsymbol{w}_\mathrm{i})$ is transformed
to
\begin{equation}
  \label{eq:transform-operation}
  (\boldsymbol{W}_\mathrm{f},\boldsymbol{w}_\mathrm{f}) =
  (\mathbf{Q}^{-1} \boldsymbol{W}_\mathrm{i} \mathbf{Q},
  \mathbf{Q}^{-1}\boldsymbol{w}_\mathrm{i}).
\end{equation}

When transforming space group operations of a primitive cell to those of a
non-primitive cell, careful consideration is required. Denote the basis vectors
of the primitive and non-primitive cells as $(\mathbf{a}_\mathrm{p},
\mathbf{b}_\mathrm{p}, \mathbf{c}_\mathrm{p})$ and $(\mathbf{a}_\mathrm{np},
\mathbf{b}_\mathrm{np}, \mathbf{c}_\mathrm{np})$, respectively. A
change-of-basis matrix relates them as
\begin{equation}
  \label{eq:non-prim-matrix}
  (\mathbf{a}_\mathrm{np}, \mathbf{b}_\mathrm{np},\mathbf{c}_\mathrm{np})
  = (\mathbf{a}_\mathrm{p}, \mathbf{b}_\mathrm{p}, \mathbf{c}_\mathrm{p})
  \mathbf{Q}.
\end{equation}
We expect
\begin{equation}
  \label{eq:non-prim-trans}
  \boldsymbol{W}_\mathrm{np}=\mathbf{Q}^{-1} \boldsymbol{W}_\mathrm{p}
  \mathbf{Q}.
\end{equation}
However, when $\mathbf{Q}$ breaks the point group symmetry of
$(\mathbf{a}_\mathrm{p}, \mathbf{b}_\mathrm{p}, \mathbf{c}_\mathrm{p})$,
$\boldsymbol{W}_\mathrm{np}$ becomes a non-integer matrix. This situation arises
when the lattice point group types associated with the basis vectors
$(\mathbf{a}_\mathrm{p}, \mathbf{b}_\mathrm{p}, \mathbf{c}_\mathrm{p})$ and
$(\mathbf{a}_\mathrm{np}, \mathbf{b}_\mathrm{np}, \mathbf{c}_\mathrm{np})$ of
the primitive and non-primitive cells, respectively, are not equivalent.

\section{\texttt{spglib} convention of symmetrization of basis vectors}
\label{sec:symmetrization-of-basis-vectors}

The \texttt{spglib} code uses specific conventions to idealize crystal
structures, which are detailed for each crystal system.

\subsection{Triclinic}

(1) Niggli reduced cell is used for choosing $\mathbf{a}$, $\mathbf{b}$, and
$\mathbf{c}$. (2) $\mathbf{a}$ is aligned with $+x$ direction of Cartesian
coordinates. (3) $\mathbf{b}$ is positioned in $x\text{-}y$ plane of Cartesian
coordinates, ensuring that $\mathbf{a}\times\mathbf{b}$ aligns with $+z$ direction
of Cartesian coordinates.

\subsection{Monoclinic}

(1) $b$ axis is taken as the unique axis. (2) $\alpha = 90^\circ$ and $\gamma =
90^\circ$ (3) $90^\circ < \beta < 120^\circ$. (4) $\mathbf{a}$ is aligned with
$+x$ direction of Cartesian coordinates. (5) $\mathbf{b}$ is aligned with $+y$
direction of Cartesian coordinates. (6) $\mathbf{c}$ is positioned in $x\text{-}z$
plane of Cartesian coordinates.

\subsection{Orthorhombic}
(1) $\alpha = \beta =
  \gamma = 90^\circ$. (2) $\mathbf{a}$ is aligned with $+x$ direction of Cartesian
coordinates. (3) $\mathbf{b}$ is aligned with $+y$ direction of Cartesian
coordinates. (4) $\mathbf{c}$ is aligned with $+z$ direction of Cartesian
coordinates.

\subsection{Tetragonal}
(1) $\alpha = \beta = \gamma =
90^\circ$. (2) $a=b$. (3) $\mathbf{a}$ is aligned with $+x$ direction of Cartesian
coordinates. (4) $\mathbf{b}$ is aligned with $+y$ direction of Cartesian
coordinates. (5) $\mathbf{c}$ is aligned with $+z$ direction of Cartesian
coordinates.

\subsection{Rhombohedral}
(1) $\alpha = \beta = \gamma$. (2) $a=b=c$. (3) When projected onto $x\text{-}y$
plane in Cartesian coordinates, $\mathbf{a}$, $\mathbf{b}$, and $\mathbf{c}$
become $\mathbf{a}_{xy}$, $\mathbf{b}_{xy}$, and $\mathbf{c}_{xy}$,
respectively, with their angles denoted as $\alpha_{xy}$, $\beta_{xy}$,
$\gamma_{xy}$. (4) The projections of $\mathbf{a}$, $\mathbf{b}$, and
$\mathbf{c}$ along $z$-axis in Cartesian coordinates are $\mathbf{a}_{z}$,
$\mathbf{b}_{z}$, and $\mathbf{c}_{z}$, respectively. (5) $\mathbf{a}_{xy}$ is
oriented along a ray rotated $30^\circ$ counter-clockwise from the $+x$
direction in Cartesian coordinates, with $\mathbf{b}_{xy}$ and $\mathbf{c}_{xy}$
positioned at angles of $120^\circ$ and $240^\circ$ counter-clockwise from
$\mathbf{a}_{xy}$, respectively. (6) $\alpha_{xy} = \beta_{xy} = \gamma_{xy} =
120^\circ$. (7) $a_{xy} = b_{xy} = c_{xy}$. (8) $a_{z} = b_{z} = c_{z}$.

\subsection{Hexagonal}
(1) $\alpha = \beta = 90^\circ$. (2) $\gamma = 120^\circ$. (3) $a=b$. (4)
  $\mathbf{a}$ is aligned with $+x$ direction of Cartesian coordinates. (5)
  $\mathbf{b}$ is positioned in $x\text{-}y$ plane of Cartesian coordinates. (6)
  $\mathbf{c}$ is aligned with $+z$ direction of Cartesian coordinates.

\subsection{Cubic}
(1) $\alpha = \beta = \gamma = 90^\circ$. (2) $a=b=c$. (3)
$\mathbf{a}$ is aligned with $+x$ direction of Cartesian coordinates. (4)
$\mathbf{b}$ is aligned with $+y$ direction of Cartesian coordinates. (5)
$\mathbf{c}$ is aligned with $+z$ direction of Cartesian coordinates.

\section{Transformation of origin shift by space group operation}
\label{sec:althernative-origin-shift}

When considering a space group operation $(\boldsymbol{W}, \boldsymbol{w})$ as a
transformation of the coordinate system, as detailed in
Appendix~\ref{sec:transformation-spg}, the new point coordinates
${\boldsymbol{x}}_\text{new}$ are related to the original point coordinates
${\boldsymbol{x}}_\text{orig}$ through
Eq.~(\ref{eq:transformation-of-coordinates}) by
\begin{align}
  \label{eq:trans-orig-new}
  \boldsymbol{x}_\text{orig} =
  (\boldsymbol{W}, \boldsymbol{w}) \boldsymbol{x}_\text{new}.
\end{align}
By definition
(\ref{eq:transformation-of-basis}),
\begin{align}
  (\mathbf{a}_\text{new},
  \mathbf{b}_\text{new},\mathbf{c}_\text{new}) = (\mathbf{a}_\text{orig},
  \mathbf{b}_\text{orig},\mathbf{c}_\text{orig}) \boldsymbol{W}.
\end{align}
Due to the space group operation $(\boldsymbol{W}, \boldsymbol{w})$,
the sets of point coordinates in the original and new coordinate systems are equal up to lattice translation,
\begin{align}
  \{\boldsymbol{x}_\text{new}\}=\{\boldsymbol{x}_\text{orig}\}
  \;(\bmod\;\mathbb{Z}).
\end{align}
Consequently, the same set of matrix representations for the elements of
$\{(\boldsymbol{W}, \boldsymbol{w})\}$ is applicable to
$\{\boldsymbol{x}_\text{new}\}$.

An origin shift from the original coordinate system is represented in
the new coordinate system as follows. Considering an origin shift
$(\boldsymbol{I}, \boldsymbol{p})$ with respect to the original coordinate
system, the point coordinates $\boldsymbol{x}_{\boldsymbol{p}}$ after this
origin shift are related to $\boldsymbol{x}_\text{orig}$ by
\begin{align}
  \label{eq:trans-orig-p}
  \boldsymbol{x}_\text{orig} =
  (\boldsymbol{I}, \boldsymbol{p})\boldsymbol{x}_{\boldsymbol{p}}.
\end{align}
From Eqs.~(\ref{eq:trans-orig-new}) and (\ref{eq:trans-orig-p}),
$\boldsymbol{x}_\text{new}$ and $\boldsymbol{x}_{\boldsymbol{p}}$ are related as
\begin{align}
  \boldsymbol{x}_\text{new} & =
  (\boldsymbol{W}, \boldsymbol{w})^{-1}
  (\boldsymbol{I}, \boldsymbol{p})\boldsymbol{x}_{\boldsymbol{p}} \nonumber \\
                            & =
  (\boldsymbol{W}^{-1}, -\boldsymbol{W}^{-1}\boldsymbol{w})
  (\boldsymbol{I}, \boldsymbol{p})\boldsymbol{x}_{\boldsymbol{p}}.
\end{align}
Viewed from the new coordinate system,
The origin shift $\boldsymbol{p}_\text{new}$ is determined by setting
$\boldsymbol{x}_{\boldsymbol{p}}=\mathbf{0}$,
\begin{align}
  \boldsymbol{p}_\text{new} =
  (\boldsymbol{W}^{-1}, -\boldsymbol{W}^{-1}\boldsymbol{w})
  \boldsymbol{p} =
  \boldsymbol{W}^{-1} (\boldsymbol{p} -\boldsymbol{w}).
\end{align}

\bibliography{spglib}

\end{document}